\documentclass{aa}
\usepackage{graphicx}
\usepackage{bm}
\usepackage{times}
\usepackage{amsmath}
\usepackage{amssymb}
\usepackage{natbib}
\usepackage{color}
\usepackage{multirow}
\bibpunct{(}{)}{;}{a}{}{,}
\graphicspath{{./fig/}{./png/}}

\newcommand{\EQ}{\begin{equation}}
\newcommand{\EE}{\end{equation}}
\newcommand{\EQA}{\begin{eqnarray}}
\newcommand{\EEA}{\end{eqnarray}}
\newcommand{\brac}[1]{\langle #1 \rangle}
\newcommand{\pd}{\partial}

\newcommand{\mean}[1]{\overline{#1}}
\newcommand{\meanv}[1]{\overline{\bm #1}}

\newcommand{\etatz}{\eta_{\rm t0}}

\newcommand{\urms}{u_{\rm rms}}

\newcommand{\urmsp}{u^\prime_{\rm rms}}

\newcommand{\Rs}{R_{\rm s}}
\newcommand{\Rm}{R_{\rm m}}
\newcommand{\Rb}{R_{\rm b}}

\newcommand{\kef}{k_{\rm f}}

\newcommand{\chiSGS}{\chi_{\rm SGS}}
\newcommand{\chiSGSm}{\chi^{\rm m}_{\rm SGS}}

\newcommand{\Pm}{{\rm Pr}_{\rm M}}
\newcommand{\Rem}{{{\rm Re}_{\rm M}}}
\newcommand{\Pra}{{\rm Pr}}
\newcommand{\PraSGS}{{\rm Pr}_{\rm SGS}}
\newcommand{\Ra}{{\rm Ra}}

\newcommand{\Rey}{{\rm Re}}
\newcommand{\Co}{{\rm Co}}

\newcommand{\nab}{\mbox{\boldmath $\nabla$} {}}

{}
\def\onethird{{\textstyle{1\over3}}}
\def\onehalf{{\textstyle{1\over2}}}

\newcommand{\Fig}[1]{Fig.~\ref{#1}}
\newcommand{\Figs}[2]{Figs.~\ref{#1} and \ref{#2}}

\newcommand{\s}{\,{\rm s}}
\newcommand{\m}{\,{\rm m}}
\newcommand{\yrs}{\,{\rm yrs}}

\begin{document}

\authorrunning{M. J. K\"apyl\"a et al.}
\titlerunning{Multiple dynamo modes and long-term solar activity variations}
\title{Multiple dynamo modes as a mechanism\\ for long-term solar activity variations}
\author{M. J. K\"apyl\"a
          \inst{1}
          \and
          P. J. K\"apyl\"a
          \inst{1,2,3}
          \and
          N. Olspert
          \inst{1}
          \and 
          A. Brandenburg
          \inst{3,4,5,6}
          \and
          J. Warnecke
          \inst{7,1}
          \and
          B. B. Karak
          \inst{3,7}
          \and
          J. Pelt
          \inst{8,1}
          }

   \institute{ReSoLVE Centre of Excellence, Department of Computer
     Science, Aalto University, PO Box 15400, FI-00076 Aalto,
     Finland\\
     \email{maarit.kapyla@aalto.fi}
         \and Department of Physics, Gustaf H\"allstr\"omin katu 2a (PO Box 64), FI-00014 University of Helsinki, Finland
         \and Nordita, KTH Royal Institute of Technology and Stockholm University, Roslagstullsbacken 23, SE-10691 Stockholm, Sweden
         \and Department of Astronomy, AlbaNova University Center,
              Stockholm University, SE-10691 Stockholm, Sweden
         \and JILA and Department of Astrophysical and Planetary Sciences,
              Box 440, University of Colorado, Boulder, CO 80303, USA
         \and Laboratory for Atmospheric and Space Physics,
              3665 Discovery Drive, Boulder, CO 80303, USA
         \and Max-Planck-Institut f\"ur Sonnensystemforschung,
         Justus-von-Liebig-Weg 3, D-37077 G\"ottingen, Germany
         \and Tartu Observatory, T\~oravere, EE-61602, Estonia
         }

   \date{Received ? / Accepted ?}

   \abstract{Solar magnetic activity shows both smooth secular
     changes, such as the Grand Modern Maximum, and quite abrupt    
     drops that are denoted as grand minima, such as the Maunder
     Minimum.
     Direct numerical
     simulations (DNS) of convection-driven dynamos offer one way of
     examining the mechanisms behind these events.}
   {In this work, we analyze a solution of a solar-like DNS that has been
     evolved for roughly 80 magnetic cycles of 4.9 years, during which epochs of
     irregular behavior are detected.
     The emphasis of our analysis is to find physical causes for such behavior.
   }
   {The DNS employed is a semi-global (wedge) magnetoconvection
     model. For the data analysis tasks we use Ensemble Empirical Mode
     Decomposition and phase dispersion methods, as they
     are well suited for analyzing cyclic (not periodic) signals.
   }
   {A special property of the DNS is the existence of multiple
     dynamo modes at different depths and latitudes.
     The dominant mode is solar-like (equatorward migration
     at low and poleward at high latitudes). This mode is accompanied by
     a higher frequency mode near the surface and at low latitudes, showing
     poleward migration, and a low-frequency mode at the bottom of the
     convection zone. The low-frequency mode is almost purely antisymmetric
     with respect to the equator, while the dominant mode has strongly
     fluctuating mixed parity. 
     The overall behavior of the dynamo solution is extremely complex
     exhibiting variable cycle lengths, epochs of disturbed and even
     ceased surface activity, and strong short-term hemispherical
     asymmetries.
     Surprisingly, the most prominent suppressed surface activity epoch
     is actually a global magnetic energy maximum, as during it the
     bottom toroidal magnetic field obtains a maximum,
     demonstrating that the interpretation of grand minima-type
     events is nontrivial. The hemispherical asymmetries are seen
     only in the magnetic field, while the velocity field exhibits
     considerably weaker asymmetry.
   }
   {
     We interpret the overall irregular behavior to be due to the interplay
     of the different dynamo modes
     showing different equatorial symmetries, especially the smoother
     part of the irregular variations being related to the variations
     of the mode strengths, evolving with different and variable cycle lengths.
     The abrupt low activity epoch in the dominant dynamo mode
     near the surface
     is related to a strong maximum of the bottom toroidal field strength,
     which causes abrupt disturbances especially in the differential rotation
     profile via the suppression of the Reynolds stresses.
   }%


   \maketitle


\section{Introduction}

Solar activity manifests itself through the well-known 11-year sunspot
cycle, during which sunspots are formed progressively closer to the
equator.
This is best seen in the time-latitude domain resulting in the so-called
butterfly diagram.
The cycle is not strictly periodic: both cycle length
and amplitude are known to be variable over time. While there are
signs of a long-term cyclic modulation on a centennial timescale referred
to as the Gleissberg cycle, also frequent grand minima with very low
activity indicators are known in solar history. 
The Maunder minimum (1645--1715) and the Dalton minimum (1790--1830)
are two prime examples of such minima in existing historical sunspot records.
Several such events have been retrieved also indirectly from
cosmogenic radionucleid data over several millennia
\citep[e.g.][]{USK07}. Especially the actual
duration and level of activity of the Maunder minimum (hereafter
MM) still continues to raise debate. For example, 
\citet{ZP15} claimed to have found historical evidence
showing that some observers did not mark down all the
sunspot observations on purpose due to the influence of religious or
philosophical dogmas, resulting in the underestimation of spottedness
during the MM, which, according to their
interpretation, was actually rather typical cyclic activity during a
regular minimum of the centennial Gleissberg cycle. In the light of
all other available activity indicators (auroral sightings, cosmogenic
radionuclide data, solar eclipse observations), analyzed by
\citet{Usoskinetal15}, this interpretation seems however unlikely.
Moreover, during the MM, the latitude range where sunspots appeared
(i.e.\ the width of the butterfly wings) was narrower
\citep[e.g.][]{RNR93,IM11,Usoskinetal15} and strong hemispherical
asymmetry was present \citep[e.g.][]{RNR93,SNR94}. Analysis of sunspot
proper motion seem to indicate slower rotation and stronger
latitudinal surface differential
rotation during the MM than for more prominent activity states
\citep{EGT76,RNR93}.

The solar magnetic field is thought to arise as an interplay of
rotation, non-uniformities related to it, and the collective inductive
effects of small-scale convective turbulent motions that amplify and
sustain the magnetic field against intense turbulent mixing
\citep[see, e.g.,][and references therein]{O03,Ch14}. The
classical hydromagnetic dynamo picture relies on significant turbulence
effects throughout the convection zone, described by
tensorial turbulent
transport coefficients describing, e.g., the $\alpha$ effect, turbulent
pumping, and turbulent diffusion
\citep[e.g.][]{M78,KR80}. Mean-field models of these so-called
distributed dynamos usually employ subsets of turbulent transport
coefficients that are either analytically derived \citep[e.g.][]{PS09}
or extracted from local convection simulations \citep{KKT06}. The
other dynamo paradigm is the so-called
flux-transport scenario which relies on the existence of
highly localized field generation
regions. One such region is at the bottom of the convection zone in
the tachocline, and the other at the top. In the top layer the twist of the
buoyantly rising flux tubes due to the Coriolis force generates poloidal
field from the underlying toroidal field (the Babcock-Leighton mechanism). The two regions are
connected through turbulent diffusion and meridional flow acting as a
conveyor belt \citep[e.g.][]{CSD95,DC99,KJMCC14}.
Both of these scenarios are capable of explaining the regular
part of the solar cycle, manifested, e.g., by the equatorial symmetry
properties and the migration direction of surface magnetic fields,
provided the turbulence effects are suitably parameterized \citep[e.g.][]{Ch10}.

To explain the grand minima-type events with mean-field dynamo models,
one usually invokes fluctuations in the main dynamo drivers, i.e.,
differential rotation, meridional circulation, and the small-scale
turbulence effects.
By including such fluctuations, dynamo models are 
usually found to exhibit irregular behavior reminiscent of grand
minima-type events \citep[e.g.][]{OHS96,Moss08,CK12}. The direct
observational knowledge of the change of these quantities during a
grand minimum-type event, however, is very limited, and therefore the
mean-field modeling approach is problematic.
Another possibility is to
seek answers from direct numerical simulations of turbulent convection
either in local \citep[e.g.][]{KMB13,MS14} or global domains
\citep[e.g.][]{GCS10,KMB12,ABMT14,FF14,MMK15,SKB15}.
This is particularly promising in the
latter case, where it is possible to directly track the change of
all relevant dynamo drivers, provided that a desired type of dynamo
solution is found.
Important exceptions are the turbulent quantities,
namely the inductive and diffusive parts of the mean turbulent
electromotive force that require special techniques to be properly
separated. One such method is the so-called test-field method
\citep{SRSRC05,SRSRC07,SPD12}, the proper application of which to
solar-like global,
spherical magnetoconvection solutions is under way
\citep{WRKKB16}.

Oscillatory dynamo solutions from global magnetoconvection models have
been known for a long time \citep{Gi83,Gl85}. The first solar-like
solutions,
however, have been obtained only recently \citep{KMB12,ABMT14}, and
only a couple of
them have previously been run up to time scales of
interest for detecting the irregular variations: the EULAG
code Millennium simulation at solar rotation \citep[][hereafter EULAG-MHD]{PC14,NCP14} covers
roughly 20 magnetic cycles of 40 years half-cycle length, while the
ASH code simulation
\citep[][hereafter ABMT]{ABMT14}, 
rotating at three times the solar rate, covers roughly 24 cycles with
a cycle length of 6.2 years. The former does not exhibit significant
irregular behavior and only very weak latitudinal migration of the
magnetic field, while the latter produces a clearer equatorward
migrating branch at lower latitudes, and a poleward migrating branch
at higher latitudes, especially pronounced in the radial field.
In addition, ABMT report a particular grand minimum-type event,
during which magnetic
activity is suppressed in the equatorial surface region and
polarity reversals are not seen in the polar surface regions for roughly 5
half cycles.
The interpretation of the origin of an oscillatory
magnetic field and equatorward migration vary considerably: from a
turbulent dynamo picture producing an $\alpha \Omega$ oscillatory
solution exhibiting a latitudinal dynamo wave \citep{KMB12,WKKB14} to
the magnetic field feeding back to differential rotation via the
Lorentz-force, the field migration being related to the variations in
the differential rotation and possibly to a non-linear dynamo
wave \citep{ABMT14}.

In this paper we analyze a semi-global (wedge-shaped)
magnetoconvection simulation similar to those reported earlier
\citep{KMB12,WKKB14} with slightly varied parameters, but integrated
over a much longer time span. The obtained simulation shows solar-like
migration patterns of the toroidal magnetic field, and exhibits a
dominant cyclic dynamo mode with an average cycle length of roughly 5
years. The simulation covers roughly 80 such cycles. In addition to
this `basic' mode, two other significant modes are detected and
characterized with suitable time series analysis tools designed for
non-periodic signals: Ensemble Empirical Mode Decomposition
\citep{WU09} and phase dispersion statistics \citep{Pelt83}.
In the following, we refer to these methods as EEMD and $D^2$ statistics,
respectively. One of the main goals of this paper is to investigate
the significance of these multiple dynamo modes to the global
dynamics of the system. The solution also exhibits several epochs of
abrupt irregular behavior (disappearance of the surface activity,
sudden switches of parity, sudden changes in cycle length and
migration direction of the toroidal field),
and a physical explanation for these events are sought for by
computing proxies for the different dynamo drivers, namely
differential rotation, meridional circulation and the $\alpha$ effect,
during different activity states.

\section{The Model} \label{sect:model}

Our magnetohydrodynamic (MHD) model has been described in many earlier
studies, in particular in \cite{KMCWB13}, and the details of it will
not be repeated here.  We perform computations in a spherical wedge
using spherical polar coordinates $(r,\theta,\phi)$ corresponding to
radius, colatitude, and longitude, respectively. The computational
domain spans from $r_0=0.7\,R_\odot$ to $r_1=R_\odot$ in the
radial direction, where $R_\odot=7\cdot10^8$m is the solar radius,
from $\theta_0=\pi/12$ to $\theta_1=11\pi/12$ in colatitude
($\pm75^\circ$ latitude), and $90^\circ$ in longitude,
i.e.\ a quarter of a full sphere.
Our setup is semi-global in the sense that we exclude the polar regions.
Including the poles would require prohibitively short
timesteps. Within this domain, we solve the standard compressible
magnetohydrodynamic equations for logarithmic density $\ln \rho$,
specific entropy $s$, velocity $\bm{u}$, and magnetic vector potential
$\bm{A}$, which gives the magnetic field as
$\bm{B}=\nabla\times\bm{A}$. To close the system of equations, we
assume that the fluid obeys the ideal gas law with $p=(\gamma-1)\rho
e$, where $\gamma=c_{\rm P}/c_{\rm V}=5/3$ is the ratio of specific
heats at constant pressure and volume, respectively, and $e=c_{\rm V}
T$ is the specific internal energy, where $T$ is temperature.
The fluid is subject to
gravitational acceleration, $\bm{g}=-GM_\odot\bm{r}/r^3$, where
$G=6.67\cdot10^{-11}$ m$^3$~kg$^{-1}$~s$^{-2}$ is the gravitational
constant and $M_\odot=2.0\cdot10^{30}$ kg is the mass of the Sun, and
to rotation, the rotation vector being
$\bm\Omega_0=(\cos\theta,-\sin\theta,0)\Omega_0$. We neglect
self-gravity of the gas in the convection zone.

As explained in detail in \cite{KMCWB13}, since we cannot
get anywhere near the high Rayleigh numbers of real stars, we must
use higher diffusivities.
In the present model this implies a roughly $10^6$ times higher
luminosity in the model in comparison to the Sun. 
This allows us to reach the Kelvin-Helmholtz time scale in our simulations,
implying that our runs are thermally relaxed.
As the convective
energy flux scales as $F_{\rm conv} \sim \rho u^3$, the convective
velocity $u$ is roughly 100 times greater in the simulations than in
the Sun.
To obtain the same rotational influence on the flow as in the Sun,
we must therefore increase $\Omega$ by the same factor.
In general, the scaling of velocity and rotation rate can be written as
\begin{equation}
  {\bm u}_{\rm sim}=L_{\rm ratio}^{1/3}{\bm u}_\odot \quad {\rm and}  \quad \Omega_{\rm sim}=L_{\rm ratio}^{1/3} \Omega_{\odot},
\end{equation}
where $L_{\rm ratio}=L_0/L_\odot$, with $L_0$ and
$L_\odot\approx3.84\cdot10^{26}$~W being the luminosities of the
model and the Sun, respectively.
In the current model, as in \cite{WKKB14}, we then renormalize our
simulation to the rotation rate
$\Omega_0\equiv5\Omega_\odot$, where
$\Omega_\odot\approx2.7\cdot10^{-6}$s$^{-1}$ is the mean solar
rotation rate, corresponding to $\Omega_\odot/2\pi=430$~nHz.

In what follows we express our results in solar
units so that, say for the velocity, we quote ${\bm u}_{\rm
  sim}/L_{\rm ratio}^{1/3}$.
The scaling used here is based on dimensional arguments.
It is supported by mixing length theory \citep{Vit53} and simulations
\citep{BCNS05,MFRT12}, and should be applicable as long as the energy transport
is not yet affected by rotation \citep[see e.g.][]{YGCD13}.
Furthermore, we assume that the density at the base of the convection
zone ($r=0.7R_\odot$) has the solar value $\rho=200$ kg~m$^{-3}$.

The simulations are performed with the {\sc Pencil
  Code}\footnote{http://github.com/pencil-code}, which uses a
high-order finite difference method for solving the compressible
equations of magnetohydrodynamics.

\subsection{Initial and boundary conditions}
\label{sec:initcond}

The initial hydrostatic state is described by a polytrope with index
$n_{\rm ad}=1.5$, i.e.\ an isentropic stratification.  Our model does
not include a stable overshoot region in the bottom of the convection
zone. The density stratification is roughly 30 in the beginning, while
it decreases to about 20 in the course of the simulation. In the final
state, the number of density scale heights covered by the model is
roughly 3.  To speed up the thermal relaxation, the initial condition
is chosen not to be in thermostatic equilibrium, but closer to the
final convecting state.  We choose the heat conduction profile $K(r)$
such that it decreases toward the surface like $r^{-15}$. This means
that a very small portion of the energy is carried by radiative
diffusion near the surface \citep{BCNS05,KKB14}.  We introduce weak
small-scale Gaussian distributed noise as velocity and magnetic field
perturbations in the initial state.

Our simulations are defined through the energy flux imposed at the bottom
boundary, $F_{\rm b}=-(K \pd T/\pd r)|_{r=r_0}$,
the values of the rotation rate $\Omega_0$, the kinematic
viscosity $\nu$, the magnetic diffusivity $\eta$, and the subgrid
scale diffusivity for the entropy in the
middle of the convection zone, $\chiSGSm=\chiSGS(r_{\rm m}=0.85\,
R_\odot)$. The radial profile of $\chiSGS$ is similar to that shown in
Fig.~1 of \cite{KMB11} with the surface value being
$\chiSGS(r_1)=6\chiSGSm$, corresponding to $6.2\cdot10^8\m^2\s^{-1}$
in physical units.
The control parameters of the simulation are quantified by the thermal
and magnetic Prandtl numbers, $\Pra_{\rm SGS}=\nu/\chiSGSm$ and
$\Pm=\nu/\eta$, respectively, and the Rayleigh number
$\Ra\!=GM_\odot(\Delta r)^4/c_{\rm P} \nu \chiSGSm R_\odot^2 (-{\rm d}
s_{\rm hs}/{\rm d}r)_{r_{\rm m}}$, where the subscript `hs' indicates a
hydrostatic non-convecting state and $\Delta r=r_1-r_0=0.3\,R_\odot$
is the depth of the convection zone.

The radial and latitudinal boundaries are assumed to be impenetrable
and stress free for the flow; see Eqs.~(8)--(9) of \cite{KMCWB13}.
The magnetic field is assumed radial on the outer
boundary at $r_1$, while the latitudinal and inner radial
boundaries are perfectly conducting; see Eqs.~(10)-(12) of \cite{KMCWB13}.
Density and specific entropy have vanishing first derivatives on the
latitudinal boundaries.
On the outer radial boundary we apply a black body condition
$\sigma T^4  = -K\nabla_r T - \chiSGS \rho T \nabla_r s$,
where $\sigma$ is a modified Stefan--Boltzmann constant, which is
chosen such that the flux at the surface carries the total luminosity
through the boundary in the initial non-convecting state.

\begin{table*}[ht!]
\begin{center}
\caption{Input and diagnostic parameters together with the kinetic and magnetic energies realized in the simulation in units of $10^5$J~m$^{-3}$.}
\begin{tabular}{ccccccccccccccc}
\hline
$\PraSGS$ & $\Pm$ & $\Rey$ & $\Rem$ & $\Co$ & $\Rem_{\rm tot}$ & $\Co_{\rm tot}$ &$E_{\rm kin}$ & $E_{\rm kin}^{\rm DR}$ & $E_{\rm kin}^{\rm MC} [10^{-3}]$ & $E_{\rm kin}^{\rm fluct}$ & $E_{\rm mag}$ &$ E_{\rm mag}^{\rm tor}$ & $E_{\rm mag}^{\rm pol}$ & $E_{\rm mag}^{\rm fluct}$ \\
\hline
1 &1 &29 &29 &9.5 &55 &5.1 &4.29 &  2.45 &  3.10 &  1.83 &  0.96 &
0.21 &  0.10 &  0.66 \\
\end{tabular}
\tablefoot{The kinetic energies are defined as the energy of the
  total flow $E_{\rm kin}=\onehalf \mean{\rho {\bm u}^2}$, differential
  rotation $E_{\rm kin}^{\rm   DR}=\onehalf \rho \mean{u}_\phi^2$,
  meridional circulation $E_{\rm   kin}^{\rm MC}=\onehalf \rho
  (\mean{u}_r^2+\mean{u}_\theta^2)$, and the fluctuating velocity
  $E_{\rm kin}^{\rm fluct}=\onehalf \mean{\rho{\bm u}^{\prime\,2}}$.
  The magnetic energies are defined as the energy of the total field
  $E_{\rm mag}=\mean{{\bm B}^2}/2\mu_0$, azimuthally averaged toroidal 
  $E_{\rm mag}^{\rm tor}=\mean{B}_\phi^2/2\mu_0$, azimuthally averaged poloidal
  $E_{\rm mag}^{\rm pol}=(\mean{B}_r^2+\mean{B}_\theta^2)/2\mu_0$, and
  the fluctuating component $E_{\rm mag}^{\rm
  fluct}=\mean{{\bm B}'^2}/2\mu_0$.
}
\label{tab:energies}
\end{center}
\end{table*}

\subsection{Diagnostic quantities}
\label{sec:diagnq}

The most important non-dimensional diagnostic quantities of our model
are the fluid and magnetic Reynolds numbers, $\Rey=\urms/\nu \kef$ and
$\Rem=\urms/\eta \kef$, where
$\urms=\sqrt{(3/2)\brac{u_r^2+u_\theta^2}_{r\theta\phi t}}$ is an estimate
of the full three-dimensional rms velocity based on the values in the
meridional plane, and the subscripts indicate averaging over $r$, $\theta$,
$\phi$, and a time interval during which the run is thermally relaxed,
and $\kef=2\pi/(r_1-r_0)\approx21 R_\odot^{-1}$ is an estimate of the
wavenumber of the largest eddies.
The azimuthal velocity component is not included in the
computation because it is dominated by the differential rotation, which
would then not characterize the level of turbulence, which is what we
are interested in.
We also use the meridional distribution of turbulent velocities
$\urmsp(r,\theta)=\sqrt{\brac{{\bm u}^{\prime\,2}}_{\phi t}}$, where
the fluctuating velocity is defined via
${\bm u}^\prime={\bm u}-\overline{\bm u}$, where the overbar
  denotes an azimuthal average, see Sect.~\ref{sect:DA} for a more
  specific definition of the averages.
  Furthermore, the rotational influence on the flow is measured by the
Coriolis number $\Co=2\Omega_0/\urms \kef$.

Our simulations are characterized by $\Rey=\Rem\approx30$,
$\PraSGS=\Pm=1$, $\Co\approx9.5$, and $\Ra\approx 1.0\cdot 10^7$. The
corresponding numbers computed for the total velocity field,
i.e.\ including the azimuthal velocity due to differential rotation,
read $\Rey_{\rm tot}=\Rem_{\rm tot}\approx55$ and $\Co_{\rm
  tot}\approx5.1$.  The most important input parameters and diagnostic
quantities are also listed in Table~\ref{tab:energies}.

This run has therefore a smaller $\PraSGS$ than the runs with
equatorward migrating magnetic fields of \cite{KMB12} and
\cite{WKKB14,WKKB15},
but twice the values of $\PraSGS$ and $\Pm$ than the run with poleward
migrating field of \cite{WKKB14} by leaving the other parameters the same.
The simulation of \cite{ABMT14} has 2--4 times smaller
Prandtl numbers, but otherwise comparable parameters.
The EULAG-Millennium simulation setup is similar to that described in
\cite{GCS10} and \cite{RCGBS11}, but with some explicit diffusion added for
stability reasons. The estimated values of their $\Rey$ and $\Rem$
are in the range 30--60, and the magnetic Prandtl number is of the
order of unity.

\section{Data analysis} \label{sect:DA}

Because MHD processes are nonlinear and the resulting data
non-stationary, it is necessary to choose analysis tools that
accurately describe its cyclic components locally and adaptively. For
example, the solar cycle is known to be of varying length and
amplitude. While Fourier analysis is the most common data analysis
technique used to extract periodicities from harmonic signals, it
requires constant amplitudes and phases and is not well-suited to the
problem \citep{Barnhart11}. In this work we have chosen to utilize the
EEMD and $D^2$ statistics, both of which are suited for the analysis
of non-stationary signals. They are presented in
Appendix~\ref{subsect:EMD} and \ref{subsect:D2}, respectively.

We define mean quantities as azimuthal averages (over the $\phi$-coordinate) and
denote them by overbars. Fluctuations about the mean are denoted by a
prime, e.g.\ ${\bm B}'={\bm B} - \meanv{B}$.
We also often average the data in
time over the period of the simulations where thermal energy and
differential rotation have reached statistically saturated states.
To study the
hemispherical asymmetries, we
analyze azimuthally averaged quantities separately for the
northern and southern hemispheres.
We also often employ the toroidal vs.\ poloidal decomposition for the
azimuthally averaged, and therefore axisymmetric, magnetic field, i.e.\ write
\begin{equation}\label{torpol}
\meanv{B}=\meanv{B}_{\rm tor}+ \meanv{B}_{\rm pol},
\end{equation}
with $\meanv{B}_{\rm tor}=\mean{B}_{\phi} \hat{\bm e}_\phi$ where
$\hat{\bm e}_{\phi}$ is the unit vector in the azimuthal direction,
and where the radial and latitudinal components form the poloidal
component $\meanv{B}_{\rm pol}=(\mean{B}_r,\mean{B}_\theta,0)$.  A
similar decomposition is used for the velocity field,
$\meanv{u}=\mean{u}_\phi\hat{\bm e}_{\phi} + \meanv{u}_{\rm
    mer}$, where $\mean{u}_\phi\hat{\bm e}_{\phi}$, and
  $\meanv{u}_{\rm mer}=(\mean{u}_r,\mean{u}_\theta,0)$. The three
main depths at which we plot/analyze the time series are near the
surface, $\Rs=0.98\,R_\odot$, at the middle, $\Rm=0.85\,R_\odot$, and
close to the bottom, $\Rb=0.72\,R_\odot$, of the convection zone.

\begin{figure*}[t!]
\begin{center}
  \includegraphics[width=\textwidth]{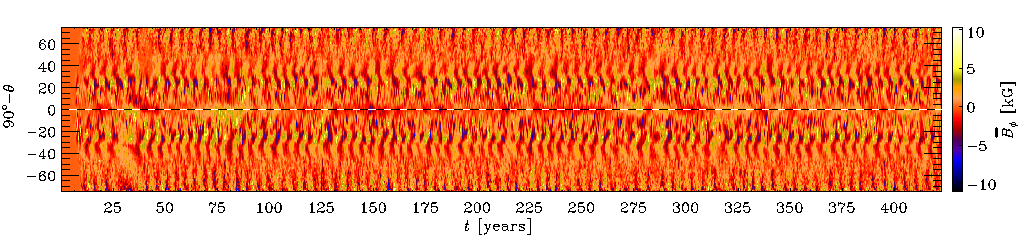}
  \includegraphics[width=\textwidth]{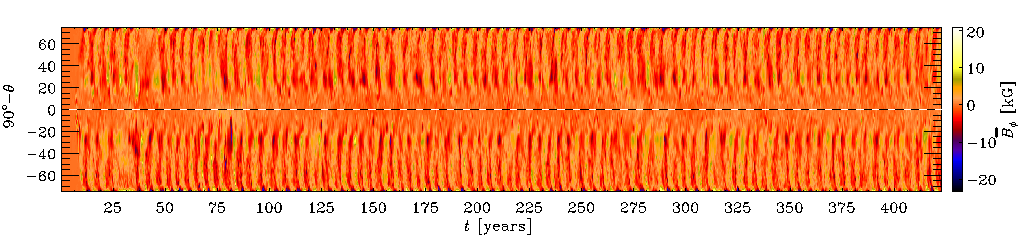}
  \includegraphics[width=\textwidth]{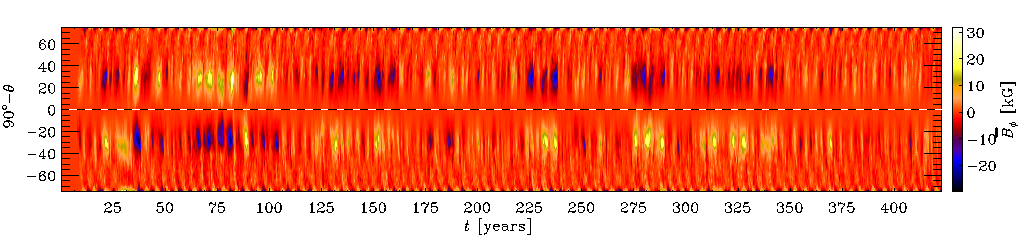}
\end{center}\caption[]{
  Time evolution of the mean toroidal magnetic field $\mean{B}_\phi$
  in the convection zone for three depths (from top to
  bottom, $\Rs$, $\Rm$, and $\Rb$). The extrema are
  $[-11.0,10.3]$\,kG at $\Rs$, $[-23.3,20.7]$\,kG at $\Rm$, and
  $[-29.9,31.7]$\,kG at $\Rb$.}
\label{buttor}
\end{figure*}

\begin{figure*}[t!]
\begin{center}
  \includegraphics[width=\textwidth]{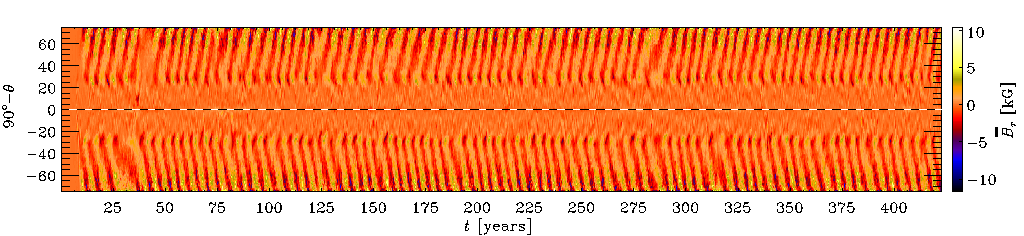}\\
  \includegraphics[width=\textwidth]{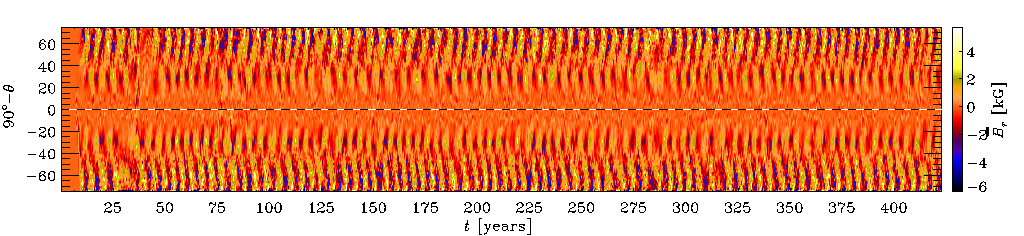}\\
  \includegraphics[width=\textwidth]{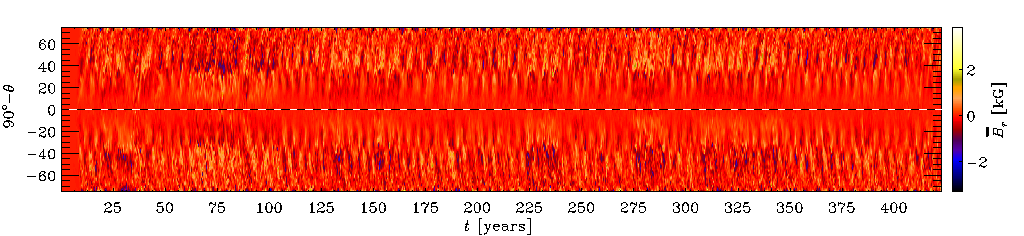}
\end{center}\caption[]{
  Evolution of the mean radial magnetic field $\mean{B}_{\rm r}$ in
  the convection zone for three depths (from top to bottom,
  $\Rs$, $\Rm$, and $\Rb$). The contours
  have been clipped at half of the maxima/minima, that is
  $[-11.7,10.4]$\,kG at $\Rs$, $[-6.1,5.7]$\,kG at $\Rm$, and
  $[-3.3,3.8]$\,kG at $\Rb$, to make the structures elsewhere than at
  high latitudes visible.}
  \label{butbr}
\end{figure*}

\section{Results} \label{sect:results}

We have run our simulation for nearly 430 years in
physical time units. The solution we obtain is oscillatory, but it shows
very complex behavior. The definition of a mean cycle period is not
unique, but, based on the near-surface toroidal magnetic field oscillation,
the mean cycle length is, in physical units, 4.9 years, 
and therefore, on average, our simulation covers 80
cycles.
In comparison to the Sun, the obtained
cycle length is roughly by a factor of 4 shorter. If, however, this was
expressed in solar units, the length of our simulation would roughly
correspond to an evolution of two millennia.
From a similar simulation with equatorward migrating fields of
\cite{WKKB15}, we find that the cycle length increases with slower
rotation, showing that with more realistic parameters it is possible
to obtain cycle lengths comparable to the solar value.
To evolve a simulation with a longer cycle over the time span presented here,
however, is computationally prohibitively expensive.
In comparison to EULAG-MHD, the number of cycles
covered here is roughly twice as large, and four times larger
in comparison to ABMT.
Furthermore, more pronounced solar-like latitudinal migration, 
and irregular behavior are observed in our simulation than
in EULAG-MHD.
In the ABMT
model, a clear solar-like equatorward migration is seen with
one MM-type event.
Some fundamental differences also
relate to the data analysis, especially concerning the EULAG-MHD: In
\citet{PC14} the analogue to the sunspot cycle is defined using the
toroidal magnetic field strength at the bottom of the convection zone,
and a poloidal field proxy is built using the radial field in the
near-surface polar regions,
while our analysis covers the whole convection zone. Moreover, we
restrain ourselves from a detailed comparison to the observed
statistical properties of the solar cycle, as we deem such attempts as
somewhat pointless due to the vast difference between simulated
and real parameter regimes, but
concentrate on the general effects that may cause the irregularities
that are seen to occur in our simulation.

\subsection{Overall evolution of the magnetic field} \label{sect:OverallM}

In Fig.~\ref{buttor} we show longitudinal averages of the toroidal
magnetic field as a function of time and latitude (butterfly diagram)
for three depths in the convection zone ($\Rs$, $\Rm$, and $\Rb$), and
the corresponding plot for the radial field in Fig.~\ref{butbr}. The
dynamo solution is, in general, very similar to those reported by
\citet{KMB12} and \cite{WKKB14,WKKB15} with cyclic behavior consisting
of an equatorward-migrating dynamo wave at low latitudes and a polar
branch at high latitudes near the surface.  The dominating component
near the surface is the radial one with roughly twice the strength of
the azimuthal component, while the azimuthal field becomes dominant at
larger depths, being roughly twice (four) times stronger than the
radial field in the middle (at the bottom) of the convection zone, see
also Figure 5(e) of \cite{WKKB14} for a plot of a similar run.  The
lack of strong toroidal magnetic field in the top layer is also
related to radial field boundary condition at the surface as
investigated in details in \cite{WKKB15}.  The marked difference to
our earlier results is the irregularity and hemispheric asymmetry seen
in the field evolution. During some epochs, the surface activity
ceases and/or becomes disturbed (i.e. the cycle length is changing
significantly), which can happen asynchronously in different
hemispheres.  The radial field evolution is more regular than that of
the toroidal field, even though the ceased surface activity is perhaps
even better seen in the radial field evolution, see
\Figs{buttor}{butbr}. During the most severe event (20-40 years in the
south; 35-45 year in the north), both the radial and toroidal fields
near the surface practically vanish. Simultaneously, regions of strong
toroidal magnetic field are seen in the near-equator regions (at
$\Rs$), and also in the bottom layers (at $\Rb$). Even visually, the
toroidal field at $\Rb$ exhibits another toroidal dynamo mode with
considerably longer period than the `basic' cycle of 4.9 years seen
throughout the simulation. In addition to these dynamo modes, there
appears to be a high frequency (short period) poleward dynamo mode in
the upper part of the convection zone, confined to the near-equator
region, similar as in \cite{KMB12} and \cite{WKKB14,WKKB15}.  This
dynamo mode might be related to a locally operating $\alpha^2$ dynamo
in the top layers \citep{KMCWB13,WKKB14}.  The quantitative
characterization of these modes is elaborated upon in
Sect.~\ref{EMDRES}.

\begin{figure}[t!]
\begin{center}
  \includegraphics[width=\columnwidth]{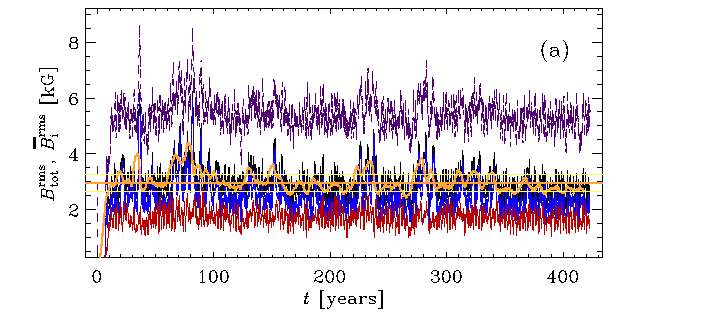}\\
  \includegraphics[width=\columnwidth]{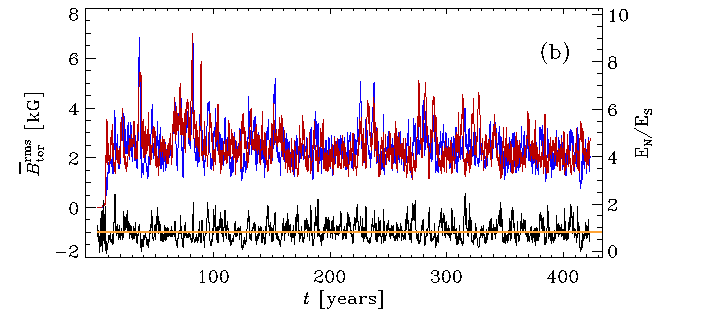}\\
  \includegraphics[width=\columnwidth]{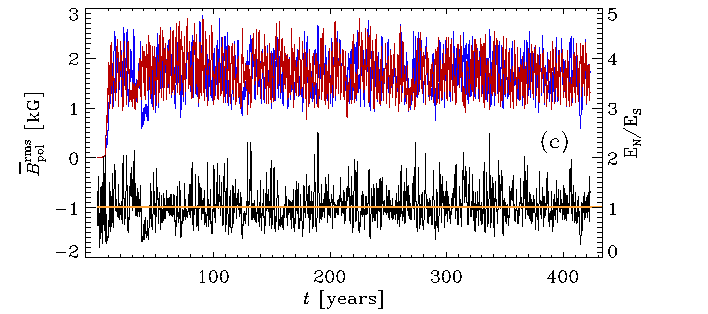}\\
\includegraphics[width=\columnwidth]{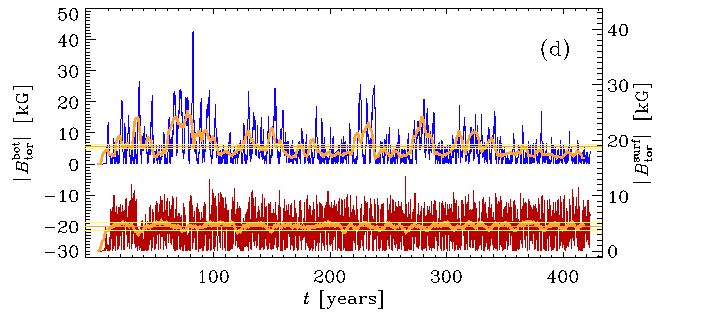}
\end{center}\caption[]{
  (a) Total ($B_{\rm tot}^{\rm rms}$ (purple dashed), total mean ($\mean{B}_{\rm tot}^{\rm rms}$, black solid), toroidal
  ($\mean{B}_{\rm tor}^{\rm rms}$, blue), and poloidal ($\mean{B}_{\rm pol}^{\rm rms}$, red)
  rms values as functions of time as averages over the whole volume.
  (b) $\mean{B}_{\rm tor}^{\rm rms}$ in the north
  (blue) and south (red). The black solid line shows the ratio of the
  northern to southern energy as function of time, and the orange
  line the mean of the ratio computed over the whole time span.
  (c) The same for the poloidal
  field.
   (d) The bottom (at $\Rb$, blue line) and surface (at $\Rs$, red line) toroidal magnetic field
  strengths as functions of time
  The thin orange horizontal lines indicate the
  mean of the quantities over the whole simulation time.
  In (a) and (d) thick orange lines show the smoothed curve of
  the total mean magnetic field (a) and the bottom/surface toroidal
  magnetic field strength (d).
  and the yellow horizontal lines the 1.1 and 0.9 times the
  mean levels, respectively.
}
\label{tottorpol}
\end{figure}

The volume- and time-averaged energies of different flow and magnetic
field components are listed in Table~\ref{tab:energies}.  In
Fig.~\ref{tottorpol}(a), we show the volume-averaged rms values of the
total and mean magnetic field together with its mean toroidal and
poloidal components as functions of time.  The dominant component is
the fluctuating part containing roughly twice as much energy as the
mean field.  In a wedge covering only one quarter of the sphere in
longitude, the large-scale magnetic field is almost purely
axisymmetric and contains only a negligible contribution from
non-axisymmetric modes.  Of the mean fields, the dominant component is
the toroidal one, containing roughly twice the energy of the poloidal
component.  This is consistent with the conclusion of \citet{WKKB14}
that the dynamo is of $\alpha \Omega$ type.  The poloidal field
evolution shows clear cyclicity with modest long-term amplitude
variations, whereas the long-term variations of the toroidal field are
much stronger. The irregularities are stronger in the early phases of
the simulation, the lowered and disturbed surface activity in the
butterfly diagrams of \Figs{buttor}{butbr} being clearly seen as a
global maximum of the volume-integrated magnetic field energy,
especially in the toroidal field, see \Fig{tottorpol}.  The early
phases, roughly 3/4 of the simulation length, show markedly irregular
behavior, while the latter part, roughly the last 1/4, shows more
regular behavior with slightly decreased overall magnetic activity
level.

In panels (b) and (c) of Fig.~\ref{tottorpol}, we plot the rms values of
toroidal (b) and poloidal (c) components for both
hemispheres (north -- blue; south -- red). The solid black lines in
these panels show the ratio of northern to southern energy (expected
to attain the value of unity if the hemispheres are totally
symmetric). Although averaged over the full time series, the
asymmetry measure is close to unity for both the toroidal (2.7\%
dominance of the northern hemisphere) and poloidal components (1.7\%
for the north), the variance around the mean is significant (around
0.1) for both field components, although the volume-integrated
quantities indicate no clear correlation between the asymmetry and
extrema in the total energy. We will investigate the north-south-asymmetry in
more detail in Sect.~\ref{sect:NS}.

In Fig.~\ref{tottorpol}(d) we plot the toroidal magnetic
field strength near the surface (at $\Rs$, red line) and the bottom (at $\Rb$, blue line). Roughly
thrice stronger toroidal fields are located at the bottom of the
convection zone in comparison to the surface. The surface field shows
some variability, but significantly smaller than those associated with
the bottom toroidal magnetic field. While the surface
toroidal field strength shows no systematic trend when the early and
late parts of the simulation are compared, the bottom field is clearly
decaying in strength. Therefore, the overall decrease and change in
the magnitude of the irregularities in the total mean field evolution
can be attributed to the changes seen in the bottom toroidal
field.

\begin{figure}[t!]
\begin{center}
\includegraphics[width=\columnwidth]{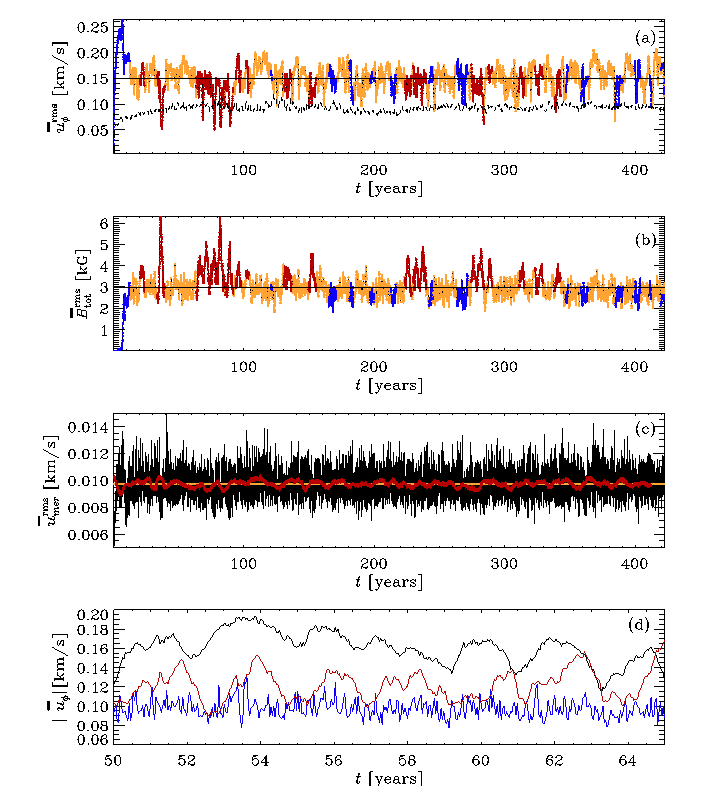}
\end{center}\caption[]{
  (a) The rms value of the mean azimuthal velocity as a function of time
  together with
  the average over the whole time series plotted with a black
  horizontal
  line.
  The dashed black line shows the volume-averaged rms value of the
  fluctuating velocity field, $\urms$.
  The red colour indicates the times which we label as `high', blue
  symbols `low', and orange `nominal' activity states according to
  the strength of the total magnetic field (D1, see Sect.~\ref{var}).
  (b) The rms value of the total magnetic field as a function of time,
  the same colour coding of overplotted symbols as in (a). The black
  horizontal line indicates the mean over the whole time series.  (c)
  The rms value of the meridional velocity as a function of time
  (black solid line) with the mean over the whole time series
  (orange horizontal line), and a 4.9-year moving average
    (thick red
  solid line).
  (d) A zoom-in to 50-65
  years of evolution of the mean azimuthal velocity (black solid),
  mean meridional velocity (blue line, multiplied by a factor of 10 to
  fit in the figure), and the total mean magnetic field strength (red
  line, scaled to fit the figure).}
\label{totU}
\end{figure}

\subsection{Overall evolution of the velocity field} \label{sect:OverallV}

In Fig.~\ref{totU} we show the rms values of the azimuthal,
fluctuating (a) and meridional (c) velocities.
The system energetics is dominated by differential rotation,
the energy of which comprises roughly 57 percent of the total kinetic
energy, being roughly 2.6 times stronger than the total magnetic energy.
The energy contained in the meridional flow is negligibly
small; in terms of the rms velocities, the meridional flow is roughly
15 times weaker than the azimuthal one.  The energy of the
fluctuations is roughly 1.3 times weaker than the energy of the
axisymmetric part.
Evidently, the azimuthal velocity becomes
strongly affected by the dynamically significant magnetic field. A
strong global magnetic field quenches the azimuthal velocity, while
during weaker magnetic field epochs, the average azimuthal velocity is
higher, as also indicated by the color coding applied in
Fig.~\ref{totU}(a) and (b) for high, low and nominal states defined
based on the global magnetic energy level (D1; see Sect.~\ref{var} for
detailed definition).  The average meridional flow is an order of
magnitude weaker, but shows signs of weak systematic variation when a
running average with the mean magnetic cycle length is applied. In
Fig.~\ref{totU}(d) we show a zoom-in over a time epoch (50--65
  years) during which no significant extrema are seen neither in the
total magnetic field strength nor the azimuthal velocity.  This figure
reveals that both the azimuthal and meridional velocities show a
variation over the magnetic cycle, stronger in the former, weaker, yet
noticeable in the latter. The azimuthal velocity and magnetic field
are phase shifted in such a way that the azimuthal velocity grows when
the magnetic field is weak, and decays when it is strong, in broad
agreement with \cite{WKMB12} and \cite{KKKBOP15}.  The meridional
velocity shows a similar trend, even though it is much more difficult
to detect as the fluctuations dominate the weak signal. Therefore, in
terms of the magnetic cycle, the angular velocity variations, often
called torsional oscillations, occur with twice the frequency of the
magnetic cycle itself (see Fig.~\ref{torsional} and Sect.~\ref{EMDRES}
for a more thorough analysis). Our results are in fair agreement with
those obtained for the EULAG-MHD and ABMT models during the regular
periods of evolution.

In contrast to the magnetic field, the mean velocity field components
show only very little hemispherical asymmetry, the evolution
of the quantities being virtually identical for both, and are therefore
not separated for north and south in Fig.~\ref{totU}. It is evident
that the strong, frequently occurring, short-term asymmetries are a
property of the magnetic field alone and any asymmetry in the flow is
probably caused by the asymmetry in the magnetic field. In the top
panel of Fig.~\ref{torsional} we plot the time-latitude diagram of the
angular velocity variations at $\Rs$. From this plot it can be seen
that on top of the rather weak, regular oscillation with twice the
magnetic cycle frequency, which is evident at all depths, there is a
rather strong long-term variation of the angular velocity near the
equator in the top layers. The origin of this variation is analyzed in
detail in Sect.~\ref{sect:drot}.

{\renewcommand{\arraystretch}{1.3}
\begin{table}\caption{Summary of the mode quantities.}
  \begin{center}
    \begin{tabular}{c|c|cccc|c|c} \hline
 \multirow{2}{*}{No.} & \multirow{2}{*}{Period} &\multicolumn{4}{c|}{Energy fraction} & {$M^{\rm reg}_i$} & \multirow{2}{*}{Eq. sym.} \\ \cline{3-6}
   & & Tot. & $\Rb$ & $\Rm$ &$\Rs$ & & \\ \hline
 \multicolumn{7}{c}{$\mean{B}_{\phi}$} \\ \hline
1  & 0.11 & -    & -    & -    & 0.13 & 0    &  0.01 \\
7  & 4.8  & 0.46 & 0.27 & 0.55 & 0.39 & 0.41 & -0.12 \\
8  & 7.0  & 0.11 & 0.16 & -    & -    & 0.15 & -0.13 \\
9  & 14   & -    & 0.11 & -    & -    & 0.09 & -0.18 \\
10 & 27   & -    & 0.12 & -    & -    & 0.18 & -0.38 \\
11 & 53   & -    & 0.11 & -    & -    & 0.35 & -0.36 \\ \hline
\multicolumn{7}{c}{$\mean{B}_r$} \\ \hline
1  & 0.11 & -    & 0.17 & -    & -    & 0    & 0 \\
7  & 5.0  & 0.62 & 0.29 & 0.60 & 0.70 & 0.75 & -0.10 \\
11 & 52   & -    & -    & -    & -    & 0.28 & -0.36 \\ \hline
\multicolumn{7}{c}{$\mean{B}_{\theta}$} \\ \hline
1  & 0.11 & -    & -    & -    & 0.25 & 0    & -0.01 \\
7  & 5.0  & 0.68 & 0.44 & 0.75 & 0.32 & 0.74 & -0.10 \\
8  & 7.4  & -    & 0.11 & -    & -    & 0.14 & -0.15 \\
11 & 62   & -    & -    & -    & -    & 0.37 & -0.58 \\ \hline\hline
\multicolumn{7}{c}{$\mean{u}_{\phi}$} \\ \hline
6  & 2.2 & 0.19 & 0.25 & 0.21 & 0.17 & 0.03 & 0.78 \\
7  & 3.7 & 0.14 & 0.11 & 0.13 & 0.15 & 0.04 & 0.75 \\
8  & 7.2 & 0.14 & -    & 0.13 & 0.14 & 0.09 & 0.73 \\
9  & 14  & 0.11 & -    & -    & 0.12 & 0.15 & 0.70 \\
10 & 26  & -    & -    & -    & 0.11 & 0.28 & 0.74 \\
11 & 48  & -    & -    & -    & -    & 0.36 & 0.64 \\ \hline
\multicolumn{7}{c}{$\mean{u}_r$} \\ \hline
1  & 0.10 & 0.61 & 0.60 & 0.60 & 0.62 & 0.06 & 0.01 \\
2  & 0.19 & 0.15 & 0.15 & 0.15 & 0.15 & 0.03 & 0.02 \\
11 & 41   & -    & -    & -    & -    & 0.23 & 0.23 \\ \hline
\multicolumn{7}{c}{$\mean{u}_{\theta}$} \\ \hline
1  & 0.10 & 0.59 & 0.60 & 0.58 & 0.57 & 0.03 & -0.05 \\
2  & 0.19 & 0.14 & 0.14 & 0.13 & 0.14 & 0.01 & 0 \\
11 & 43   & -    & -    & -    & -    & 0.29 & 0.21 \\ \hline \hline 
\multicolumn{7}{c}{$H_{\rm kin}$} \\ \hline
1  & 0.10 & 0.57 & 0.54 & 0.56 & 0.57 & 0.02 & -0.27 \\
2  & 0.19 & 0.13 & 0.16 & 0.13 & 0.14 & 0.01 & -0.22 \\
11 & 41   & -    & -    & -    & -    & 0.38 & -0.48 \\ \hline\hline
\end{tabular}
\tablefoot{
  The most prominent modes of mean magnetic and mean velocity fields
  found
  from EEMD. Columns from left to right: mode number, average mode period 
  in years, fraction of energy contained in mode (over the
  full meridional plane, over the latitude range
  at $\Rb$, $\Rm$, and $\Rs$), regularity level,
  equatorial symmetry of the modes.
  To emphasize stronger modes we have omitted all energy
  fraction values lower than a preselected level of 10\% (indicated by ``-")
}\label{mode_summary}
\end{center}
\end{table}

\subsection{Multiple dynamo modes and their significance}\label{EMDRES}

The results presented in this section have been obtained by applying
EEMD (see App.~\ref{subsect:EMD}) analysis on the azimuthally averaged
quantities using an ensemble size of 100 and Gaussian white noise with
standard deviation equal to that of the time series being
analyzed. The time series were chosen by sampling the simulation
domain in radial direction with 12 and in latitudinal direction with
18 points.  Our analysis is focused on the components of mean magnetic
field, mean velocity field and kinetic helicity, aiming to extract the
most significant modes for the given time series globally as well as
at three depths ($\Rb$, $\Rm$, and $\Rs$).  We also use the $D^2$
statistic (see App.~\ref{subsect:D2}) on some selected data sets to
validate the EEMD results, but also to investigate the coherence of
the cycles over time.

We detected on average a total number of 13 modes.  As expected, it
turned out that most of the energy was contained only in a limited
number of modes. In the following we use a mode numbering scheme such
that the modes are ordered by their average frequency from higher to
lower values (the highest frequency mode thus having an index 1).  To
express the energy contained in each mode quantitatively we will use
the notion defined as follows:
\begin{equation}\label{mode_energy}
E_i=\frac{\iint(\overline{A_i^2})ds}{\iint(\sum\limits_{i}\overline{A_i^2})ds},
\end{equation}
where $A$ is the physical quantity analyzed, $i$ is the mode
index, $\overline{A_i^2}$ is the mean square amplitude of the mode,
and summation is performed over all modes and integration either over
latitude at selected radii or over the full meridional plane.

\begin{figure*}[t!]
\begin{center}
  \includegraphics[width=\textwidth]{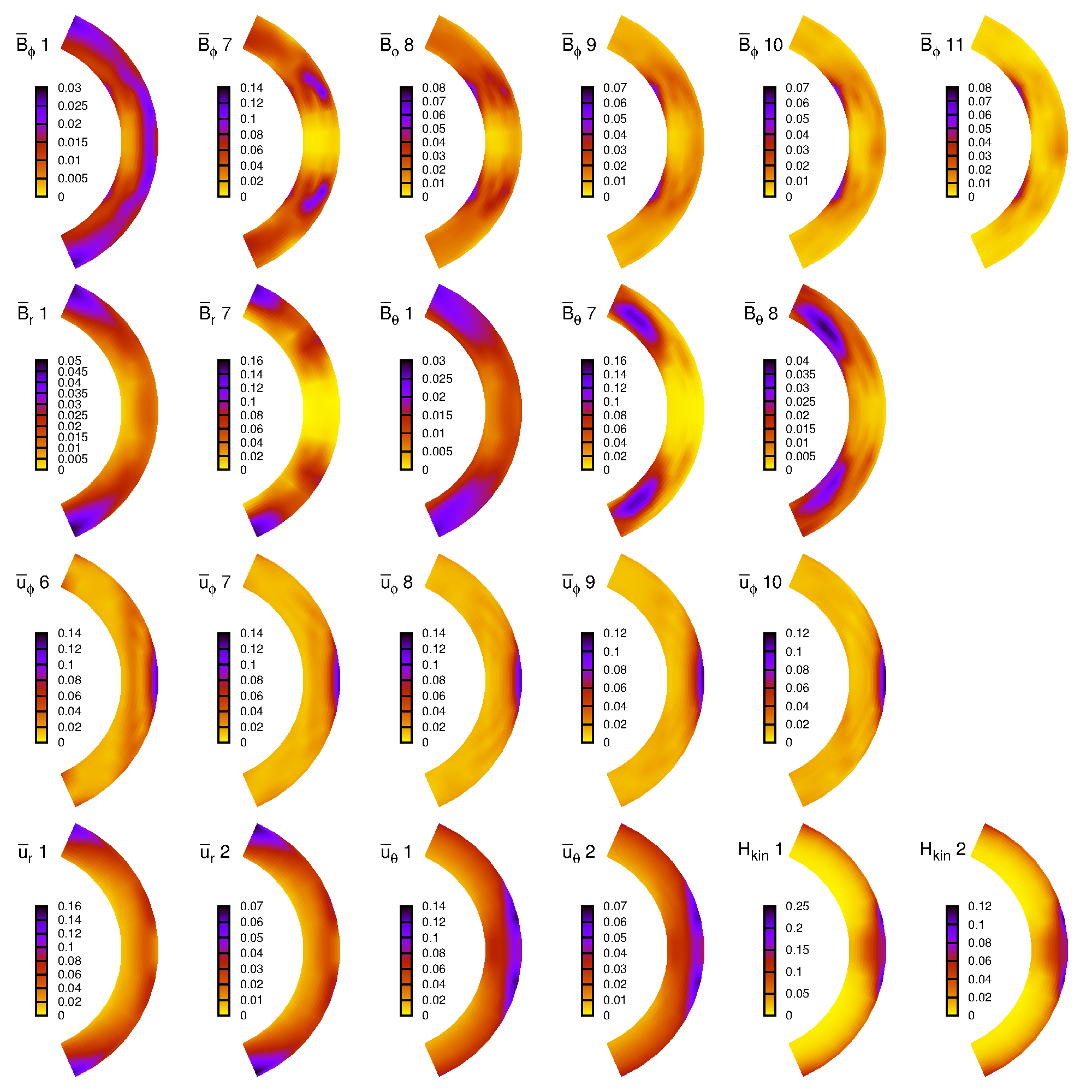}
\end{center}\caption[]{Distribution of mean amplitudes of the most
  significant modes of mean magnetic and mean velocity fields found from EEMD.
  The figures are labeled by the physical variable followed by mode index (e.g $\mean{B}_\theta$ 7
  indicates mode 7 of the mean latitudinal magnetic field). Colors
  reflect the
mean amplitude (blue -- high, yellow -- low) of the mode at the given location (we note that the scales of separate figures
  are different). Contours on the plot represent the lines of constant amplitude with values given
  in the legend.}
\label{mode_dist}
\end{figure*}

\begin{figure*}[t]
\centering
\includegraphics[width=1.0\textwidth]{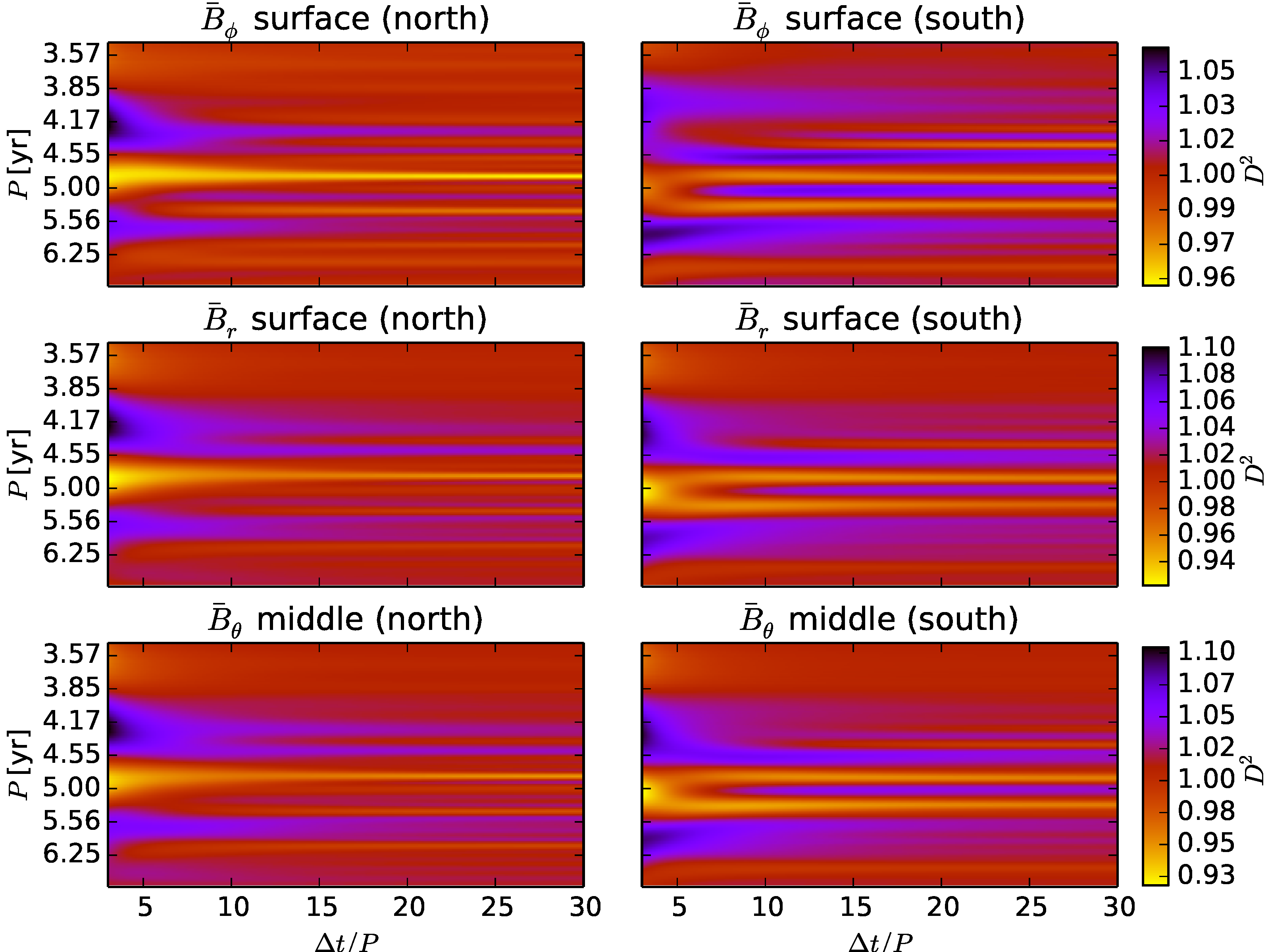}
\caption{
Phase dispersion analysis results for the components of $\meanv{B}$. The calculations were done
for $\meanv{B}_\phi$ at latitude $\pm 22^\circ$ and radius $0.94R_\odot$,
for $\meanv{B}_r$ at latitude $\pm 66^\circ$ and radius $0.94R_\odot$
and for $\meanv{B}_\theta$ at latitude $\pm 49^\circ$ and radius
$0.82R_\odot$.
Left (right) refers to the northern (southern) hemisphere.
}\label{D2}\end{figure*}

However, calculating only the energy content does not allow us to
detect weak, but more regular (harmonic) modes.  For that purpose, for
each mode we calculate the spectral entropy $H=\int p(\nu) \log p(\nu)
d\nu$, where $p(\nu)$ is the power spectral density at frequency
$\nu$. Average value of the given quantity per each mode compared to
the expected value in the case of white noise serves as a regularity
measure of the mode (i.e. how much more Gaussian or non-uniform the
spectrum is compared to the spectrum of the white noise).  An
alternative, but more straightforwardly calculated measure is the
ratio of energies in each mode to the expected ones of white noise.  A
large value of this ratio for a given mode is another indication of
the presence of a more regular signal.  This idea is captured by the
following formula:
\begin{equation} \label{mode_regularity}
M^{\rm reg}_i=\frac{E_i/E^{\rm WN}_i}{\sum\limits_{i}E_i/E^{\rm WN}_i},
\end{equation}
where $E_i$ is a mode energy fraction defined in
Eq.~(\ref{mode_energy}) and $E^{\rm WN}_i$ is the expected energy
fraction of white noise for the given mode.  As will be explained
later, the results for both regularity measures are in good agreement
with each other.  For that reason we only show in the following
quantitative values calculated based on the second definition.

The results of the EEMD analysis are gathered in
Table~\ref{mode_summary}, where the mode period represents a
descriptive period which is calculated based on the zero-crossings of
the most dominant instance of the mode of given order (i.e.\ the
intrinsic mode function, hereafter IMF, with the highest average mean
energy over the full meridional plane).  In the table we also show
fractions $E_i$ of the energy contained in the $i$th mode calculated
over the full meridional plane as well as over the latitude at radii
$\Rb$, $\Rm$ and $\Rs$.  The latter values are approximately equal to
the reconstruction error of the initial signal given that the $i$th
mode would be omitted from the sum.  For clarity reasons only the
values above 10\% are shown, lower values have been marked with ``-''.
$i$th mode would be omitted from the sum. For clarity reasons only the
values above 10\% are shown, lower values have been marked with "-". 
We can see that in the case of the magnetic field components most of
the energy is contained in mode 7 (hereafter M7), which can be
associated with the dominant cyclicity of 4.9 years cycle length
  deduced from $D^2$ statistic using surface toroidal field.  Only
for $\mean{B}_{\phi}$ in the bottom layer the energy is more spread
between modes with a longer period.  From Fig.~\ref{mode_dist}, two
upper rows showing the spatial distribution of the mean amplitudes of
the most significant modes found in the magnetic field, it is clear
that for the toroidal component of the magnetic field, M7 is mainly
confined to the region where the equatorward migration of the field is
observed (compare Fig.~\ref{mode_dist}, second panel from left
  in the upper row with Fig.~\ref{ReM}, top panel). For the poloidal
field, however, the M7 amplitude is the highest at higher latitudes:
for the radial field near the surface close to the latitudinal
boundaries, and for the latitudinal field, near the bottom and high
latitudes. The modes with shorter cycles (mode 1 being the one
containing most of the energy, hereafter M1) have the highest
amplitudes near the equatorial region and near the latitudinal
boundaries, both regions being concentrated near the surface. The long
modes (the most notable being mode 11, also having an increased
regularity measure, hereafter M11) reside at the very bottom of the
convection zone, confined mainly to low-latitude regions.

In the case of $\mean{u}_{\phi}$, the energy spectrum is more flat with roughly
equal distribution of energy between modes 6--10.
From Fig.~\ref{mode_dist}, third row, it is evident that the
highest amplitudes of these modes all occur in the equatorial region,
the region extending to higher latitudes for modes 6--7, and getting
narrower for the ones with a longer cycle.
For the other velocity components as well as for kinetic helicity, defined as 
$H_{\rm kin}=\mean{{\bm\omega}^\prime\cdot {\bm u}^\prime}$,
where ${\bm\omega}^\prime=\nab\times{\bm u}^\prime$,
energy is primarily contained only in modes 1 and 2.
From Fig.~\ref{mode_dist}, last row, we can see that these short cycles
are most prominent near the surface close to the latitudinal
boundaries for the radial component of the velocity, and near the
equator at the surface for the latitudinal velocity component. The
modes in the kinetic helicity show high amplitudes near the equatorial
region extending deeper to the convection zone, but also in a narrow
band near the surface extending the whole latitudinal extent.

In the penultimate column of Table~\ref{mode_summary}, we show a
measure of mode regularity as defined in Eq.~(\ref{mode_regularity}).
These values are calculated at those radii and latitudes where the
energy of the given mode has its maximum.  Higher values of this
number indicate higher energy content in the given mode compared to
the white noise level.  According to the values shown, besides M7 of
the magnetic field, we have one additional mode where the regularity
has clearly a high value.  This is M11, which, however, has a
significant energy fraction only in the case of $\mean{B}_{\phi}$ in
the bottom layer.  It is also quite obvious that the leading modes of
$\mean{u}_r$, $\mean{u}_{\theta}$, and $H_{\rm kin}$ represent
primarily noise, because the value of regularity in these cases is
very low.  It is noteworthy, however, that for all these quantities
the regularity measure of M11 is high, indicating the presence of a
long-term regular cycle.  Here we also note that the mean spectral
entropy calculations lead to similar results as above: the spectrum of
M7 of the magnetic field is the most regular one, while modes 10 and
11 are less regular, although significant deviations from the entropy
of the white noise spectrum are still detected.  It can be concluded,
therefore, that the dynamo mode at the bottom of the convection zone
has an influence on the overall dynamics of the system despite it
being rather insignificant in terms of the total energetics of the
system.

The last column of Table~\ref{mode_summary} represents the equatorial
symmetry of the modes, calculated as a time averaged parity defined by
Eq.~(\ref{eq:eqsym}).  The EEMD analysis confirms our earlier
conclusion that the mean azimuthal velocity field is mostly symmetric
between the hemispheres, as the parity is high. For all the magnetic
field modes, however, values significantly deviating from perfect
anti-symmetry (parity values of $-1$) or perfect symmetry (parity
values of $+1$) are found.  The higher frequency modes show parity
values fluctuating around zero, while the lower frequency modes show
increasing tendency for antisymmetric parity as the period of the
cycle increases. The equatorial symmetry will be discussed in more
detail in Sect.~\ref{sect:NS}.

As an independent check for the EEMD results we choose the $D^2$ phase
dispersion statistic (see App.~\ref{subsect:D2}).
Based on the spatial distribution of the most significant modes, we
calculate the $D^2$ spectra at selected depths for the components
of $\meanv{B}$.
Ideally, we would expect to see a
match between the average mode periods found from EEMD and cycle
lengths found from $D^2$.

To focus on the most pronounced activity regions we chose the latitude
$\pm 22^\circ$ in the layer at $\Rs$ for $\mean{B}_{\phi}$, 
latitude $\pm 66^\circ$ in the layer at $\Rs$ for $\mean{B}_r$, and 
latitude $\pm 49^\circ$ in the layer at $\Rm$ for $\mean{B}_{\theta}$
for detecting the short cycles such as M7, see Table~\ref{mode_summary}.
Due to the limited length of the data set, the $D^2$ analysis is not feasible
for detecting the long cycle.
However, by using low-pass filtering on
the raw data, the existence of M11 can be easily confirmed.
The results of the $D^2$ analysis are depicted in Fig.~\ref{D2}. 
Evidence of M7 with a cycle length of around 5 years can be seen for
all components.
It is interesting, however, that the cycle is modulated differently for northern and 
southern hemispheres.
For the southern hemisphere there is an indication of amplitude modulation
with a long period, while for the northern hemisphere
the modulation seems to be more complex.
If we assume a simple amplitude modulation for the southern hemisphere, 
and this is justified by the fact that we see a relatively symmetric split
of the spectrum when moving from short coherence times to longer ones,
then by measuring the difference of these peaks from the spectrum taken at the
high end of the coherence time, we can estimate the period of this modulation.
Given the low precision that we can
achieve using this simple procedure we obtained the value for the
modulating period to be around 100$\pm$10 years.

We also analyzed the higher frequency region (0.1 to 2 years) with the
$D^2$ statistic, but did not detect any strong and/or singular minima
in this range. Taken that the regularity measures obtained from EEMD
also indicate very low values for these modes, we have to conclude
that this cyclic mode is coherent only over very short time intervals,
the variation of the length of this cycle being comparable to the
length of the cycle itself. Therefore, the signal of this cycle can be
spread between multiple EEMD modes, and the spectrum smeared out over
a large frequency band.

\subsection{Definition of the activity levels}\label{var}

As was evident from the results presented in
Sect.~\ref{sect:OverallM}, a
decreased surface magnetic activity does not straightforwardly imply a global
magnetic energy minimum, especially during those epochs when the
long-period dynamo mode at the bottom of the convection zone is strong
(roughly speaking the first 3/4 of the simulation). Therefore,
the definition of a low/high state is not obvious. Here we
use three different definitions, all based on computing running
averages with a window coinciding with the main periodicity present
at different depths and latitudes (4.9 years):
\begin{itemize}
\item[D1.]
If the volume-averaged mean magnetic energy is larger/smaller than
the variation of its smoothed average, the state is termed high/low.
All other epochs are
  termed nominal states of activity.
\item[D2.] Otherwise identical, but the surface magnetic field energy at $\Rs$ in
  comparison to its smoothed value is used as criterion.
\item[D3.] Otherwise identical, but the energy contained in the
  magnetic field mode of the bottom layer at $\Rb$ is used as criterion.
\end{itemize}
In Fig.~\ref{tottorpol}(a; D1) and (d; D2 and D3) we demonstrate these
criteria by overplotting the smoothed curves (orange thick lines) and
the corresponding upper/lower limits (yellow horizontal lines) on top
of the original data.
In the remaining parts of this section, we use these definitions to
analyze some key quantities during the different activity states.
In practise this
is done such that we compute the mean of a quantity over the whole
simulation run, divide the data points into different states according
to each criterion, and compute the mean profiles for the three
activity states separately. If the so-obtained profiles differ significantly
from the global mean, i.e.\ the difference is larger than the
standard deviation of the quantity, we take a note of this dependence.

\begin{figure*}[t!]
\begin{center}
  \includegraphics[width=\textwidth]{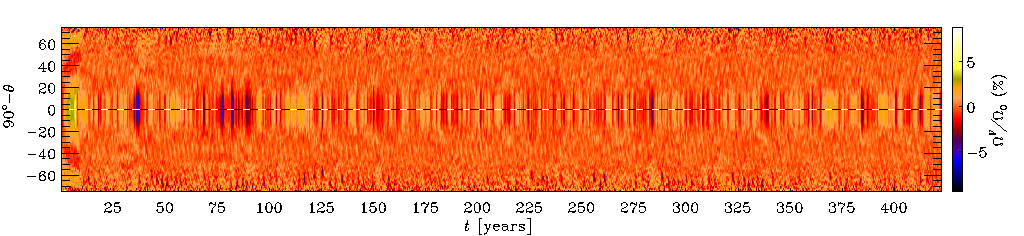}\\
  \includegraphics[width=\textwidth]{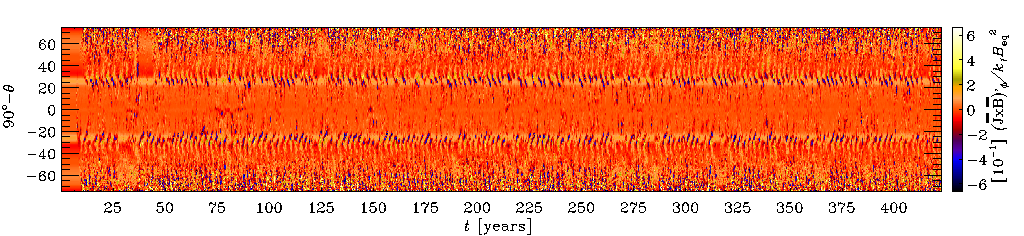}\\
  \includegraphics[width=\textwidth]{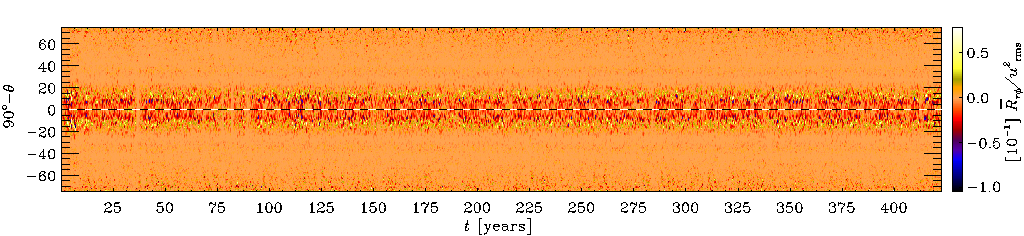}\\
  \includegraphics[width=\textwidth]{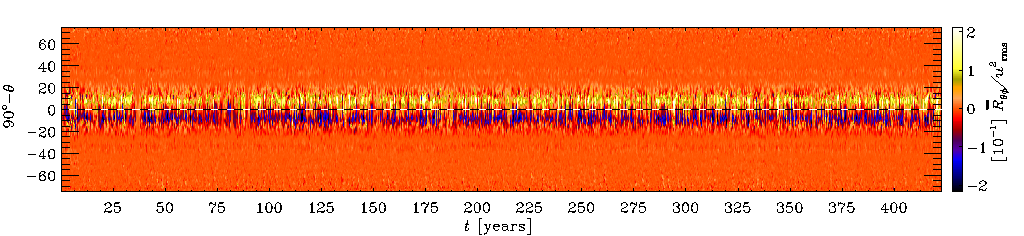}  
\end{center}\caption[]{From top to bottom: Time-latitude diagram of 
  the angular velocity variations $\Omega^{\rm v}$,
variations of the large-scale Lorentz force $\mean{\bm
  J}\times\mean{\bm B}$, and the Reynolds stress
components $R_{r \phi}$
and $R_{\theta \phi}$ near the surface (at $\Rs$).
For the Reynolds stress plots, the contours have been clipped
at half of the maxima/minima.}
\label{torsional}
\end{figure*}

\subsection{Variation of the dynamo numbers}\label{var}

In this section, we investigate the variability of the three key
factors involved in the dynamo process, namely differential rotation
and meridional circulation (Sect.~\ref{sect:drot}), as well as the
inductive effects due to convective turbulence ($\alpha$ effect;
through kinetic and magnetic helicities as its proxy,
Sect.~\ref{sect:alp}), using the different definitions of the activity
level.

\subsubsection{Rotation and its non-uniformities}\label{sect:drot}

As stated in the Sect.~\ref{sect:model}, the input angular rotation
velocity is five times the solar rotation rate. The differential
rotation profile generated in the simulation is solar-like, although
the equator is rotating only approximately 7\% faster than the poles
when the magnetic fields have grown and saturated at dynamically
significant strengths.  Therefore, the obtained latitudinal
differential rotation is approximately three times weaker than in the
Sun.  We have also performed a hydrodynamical counterpart simulation,
in which roughly twice stronger latitudinal differential rotation
(pole-equator difference of 14\%) is generated.  The distribution of
angular velocity is fairly similar to the MHD state, discussed later
in this section, and therefore the HD results are not shown
separately.  Furthermore, the hydrodynamic state is steady, and shows
no oscillatory behavior similar to the dynamo run.  Therefore, the
variations seen in the MHD state (Fig.~\ref{totU}) arise as a
consequence of the dynamically significant magnetic field to the flow
field.  Such a backreaction can occur through two pathways: a
large-scale Lorentz-force $\mean{\bm J}\times\mean{\bm B}$ \citep[the
  so-called Malkus-Proctor effect; see][]{MP75} or through turbulence
effects either directly on the convective motions and thereby the
generators of differential rotation \cite[the $\Lambda$-effect:
  see][]{R89} as recently explored in \citet{KKKBOP15}, or through the
turbulent Maxwell stress \citep[e.g.][]{KKT04}.  At the same time as
quenching the flow field, the magnetic field suppresses its own
generators.  One argument to explain both the regular and irregular
parts of the solar cycle is related to this mechanism (see
e.g.\ ABMT).  The classical theory of turbulent hydromagnetic dynamos,
however, allows for the excitation of oscillatory solutions
independent of the pre-existence of such modulation in the flow field
\citep[see e.g.][]{Pa55b}. In this section we set out to investigate
the role of the changes seen in the mean flow to the long-term
evolution of the magnetic field, especially to its disturbed states.

We begin by investigating the angular velocity variations $\Omega^{\rm
  v}$ over time in more detail.  Here $\Omega^{\rm
  v}=\Omega-\brac{\Omega}_t$, where
$\Omega=\mean{u}_{\phi}/r\sin\theta+\Omega_0$ and $\brac{\Omega}_t$ is
time-averaged over the saturated stage.  We plot a time-latitude
diagram of this quantity at $\Rs$ (Fig.~\ref{torsional}, top panel),
at which depth the most prominent variations occur in the equatorial
regions.  In addition to the strongest, seemingly irregular, changes
symmetrically distributed around the equator, a weaker modulation
corresponding to the poleward migrating magnetic cycle can be seen at
high latitudes throughout the simulation.  In Fig.~\ref{torsional},
second panel, we investigate the role of the large-scale Lorentz force
$\mean{\bm J}\times\mean{\bm B}$ in causing the changes seen in the
angular velocity. It is evident that the high-latitude variations in
the angular velocity, following the mean magnetic cycle, are mediated
through this effect. The sudden drops of the surface magnetic field
strength are very clearly seen as similar behavior in the large-scale
Lorentz force, and occur simultaneously in time in comparison to the
surface field disappearances, especially pronounced during the first
50 years of the simulation.  Only very mild enhancements of
differential rotation at higher latitudes are accompanied by the
sudden drops in the Lorentz force.  The strong variations seen in the
angular velocity near the equator, on the other hand, cannot be
explained by the large-scale Lorentz force.

Their symmetric character with respect to the equator hints towards
them being related to the Reynolds stress component $R_{r
  \phi}=\mean{u^\prime_r u^\prime_\phi}$. Instead of inspecting all
the Reynolds stress components, we concentrate here on examining the
variability of those known to be the most significant for the
generation of differential rotation, namely $R_{r \phi}$ for vertical
and $R_{\theta \phi}=\mean{u^\prime_\theta u^\prime_\phi}$ for
horizontal differential rotation, and postpone the full analysis of
the turbulent quantities to a forthcoming paper (hereafter Paper II).
The time-latitude plot of the aforementioned Reynolds stress
components are shown in the two lowermost panels of
Fig.~\ref{torsional}. As is evident from these figures, both stress
components undergo variations on the time scale of the shortest dynamo
cycle (M1), while showing no modulation by the dominant magnetic cycle
(M7). The stresses have non-zero values only near the equatorial
regions, where the strongest angular velocity variations are seen. The
minima/maxima of the stresses coincide very well with the
minima/maxima of the angular velocity and therefore indicate a strong
connection between these quantities.

Next we apply the three different definitions of high and low magnetic
activity states (D1--D3) introduced above, and seek for significant
deviations, indicative of the sensitivity of the mean flow field on
the quantity chosen. The mean flow is almost completely insensitive to
the variations of the surface magnetic activity level (D2), and not
significantly different when the bottom activity level is used
(D3). Some significant differences, however, can be detected using the
global magnetic activity level (D1) as an indicator.

Results with D1 are depicted in Fig.~\ref{drot},
where we compare the time-averaged rotation profile to the
different activity states
(low, nominal, and high) denoted and defined as
\begin{equation}
  \Delta \Omega_{\rm state}=(\brac{\Omega}_t-\brac{\Omega}_{\rm
    state})/\Omega_0,
\end{equation}
where $\brac{\Omega}_t$ is the time-averaged rotation profile,
computed over the whole simulation excluding only the initial
kinematic stage when the magnetic field is still growing, and the
quantities including angular brackets denote averages over a certain
state. Using this definition, enhancement/quenching with respect to
the average value during a certain state will show up as
negative/positive values in the plot, respectively.

The time-averaged rotation profile is very similar to the ones
obtained and reported in various earlier works \citep{KMB12,KMCWB13},
being solar-like, but with a somewhat more cylindrical profile,
showing a mid-latitude region of slower rotation leading to a region
of reversed sign of radial and latitudinal differential rotation. This
region was found to be instrumental in causing the equatorward
migration of the magnetic field in \cite{WKKB14}.  The variations in
the rotation profile averaged over different global magnetic activity
states are weak (roughly 0.4 percent), while the instantaneous
variations near the surface (see Fig.~\ref{torsional}) could be as
large as 5 percent during the early stages of the simulation.  During
a low/high state, slightly faster/slower equatorial rotation is
obtained. This is consistent with the picture that the stronger the
magnetic field, the stronger the suppression of the velocity field.
The region with a reversed gradient of $\Omega$, however, is observed
to persist during all activity states, with virtually no change in its
magnitude.

In Fig.~\ref{drot} (middle row, radial differential rotation; lower
row, latitudinal differential rotation) we investigate how the
magnitude and distribution of differential rotation change depending
on the global magnetic activity level.
Here we define the radial shear as
$\Delta_r=\partial\Omega/\partial r$
and the latitudinal shear as
$\Delta_\theta=\partial\Omega/r\partial\theta$, and the
profiles computed over difference activity states as
\begin{equation}
  \Delta_{i,\rm state}=\brac{\Delta_i}_t-\brac{\Delta}_{i,\rm state},
\end{equation}
where the quantities with angular brackets are averages over the
whole time series except for the kinematic stage (index $t$) or over a
certain state (index $i$).
It is evident that during high/low
states of magnetic activity, also the radial differential rotation is
quenched/enhanced similar to the rotation itself,
the magnitude of the change being roughly 1.5 percent in the
equatorial region.
No significant change is seen for the latitudinal differential
rotation, except for at the high latitude regions, which show hemispheric
asymmetry especially during the low state.
Finally, we define a
dynamo number describing the magnitude of the radial differential
rotation as
\begin{equation}
  C_{\rm \Omega}=\frac{\Delta_r (\Delta r)^3}{\etatz}, \label{Com}
\end{equation}
where the turbulent diffusivity is approximated by
$\etatz=\tau u^{\prime\,2}_{\rm rms}(r,\theta)/3$, with
$\tau=\alpha_{\rm MLT}H_p/\urmsp(r,\theta)$ being the local turbulent correlation time, 
$\alpha_{\rm MLT}$ the mixing length parameter and $H_p=-(\partial\ln
p(r,\theta)/\partial r)^{-1}$ the pressure scale height. Following our previous
studies, we use $\alpha_{\rm MLT}=5/3$. The spatial distribution of
$C_{\rm \Omega}$ follows closely that of the radial gradient of the
angular velocity (the leftmost panel in the middle row of
Fig.~\ref{drot}), and therefore we do not separately plot it, while
merely report its spatial extrema $[-30,110]$ averaged
  over the whole time span of the simulation, which is similar to
those obtained in \cite{KMCWB13} and \cite{WKKB14}.

\begin{figure*}[t!]
\begin{center}
  \includegraphics[width=0.18\textwidth]{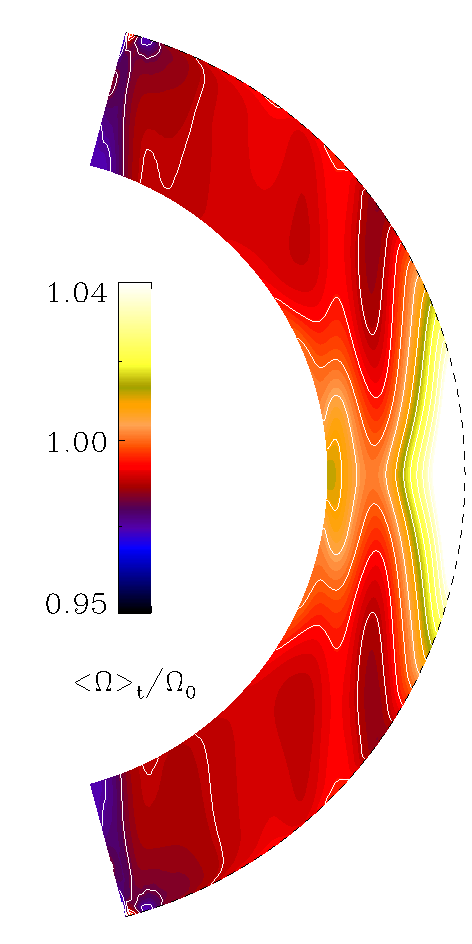}
  \includegraphics[width=0.18\textwidth]{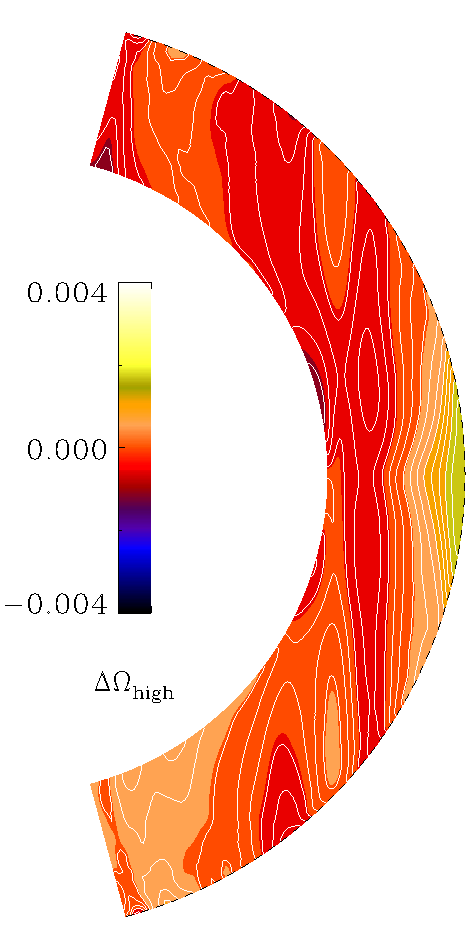}
  \includegraphics[width=0.18\textwidth]{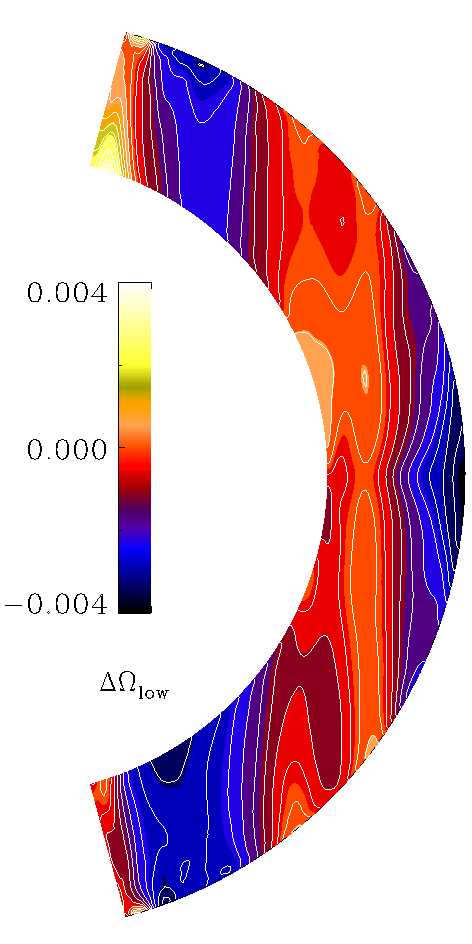}
  \includegraphics[width=0.18\textwidth]{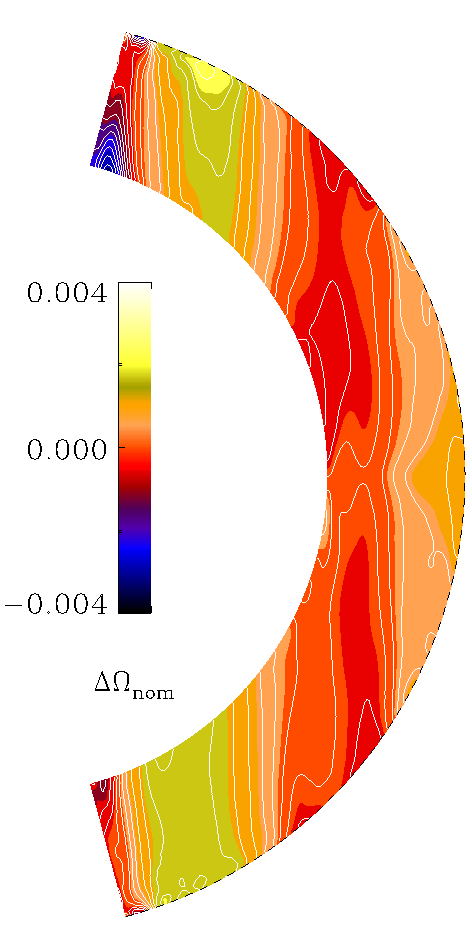}\\
    \includegraphics[width=0.18\textwidth]{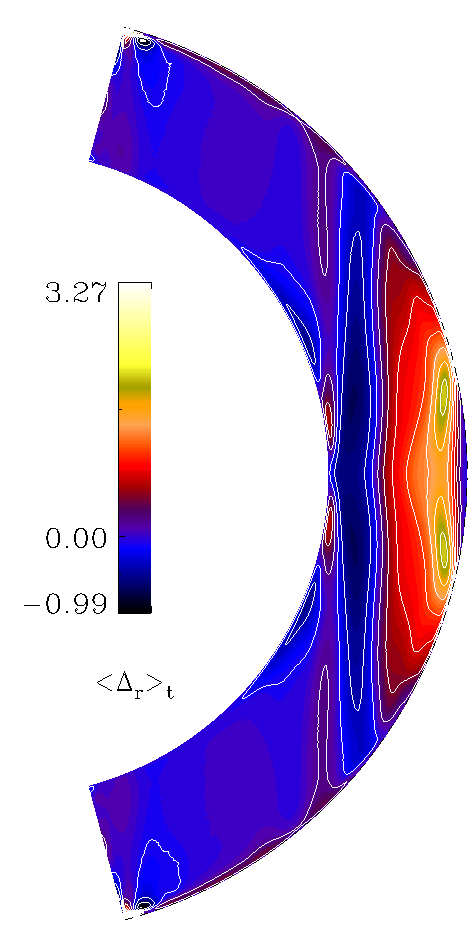}
  \includegraphics[width=0.18\textwidth]{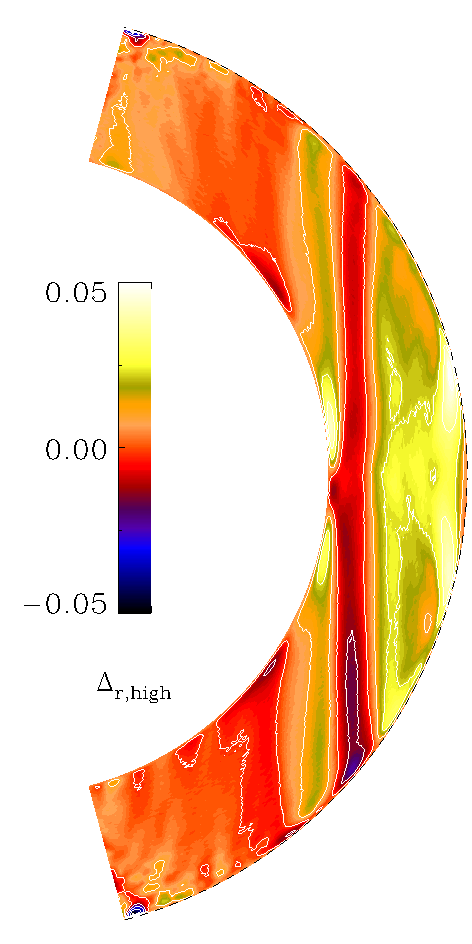}
  \includegraphics[width=0.18\textwidth]{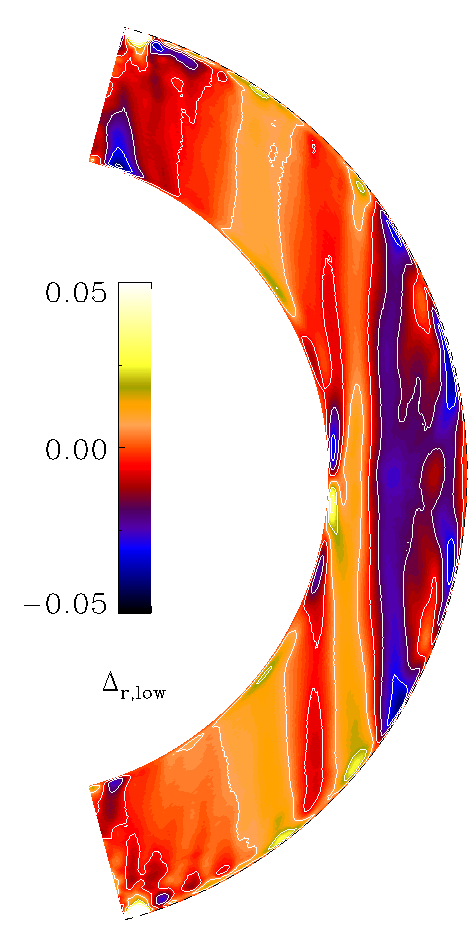}
  \includegraphics[width=0.18\textwidth]{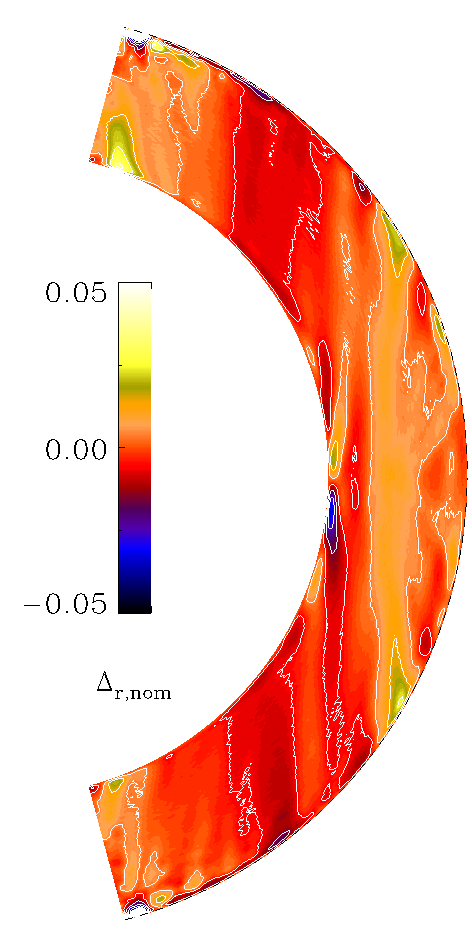}\\
      \includegraphics[width=0.18\textwidth]{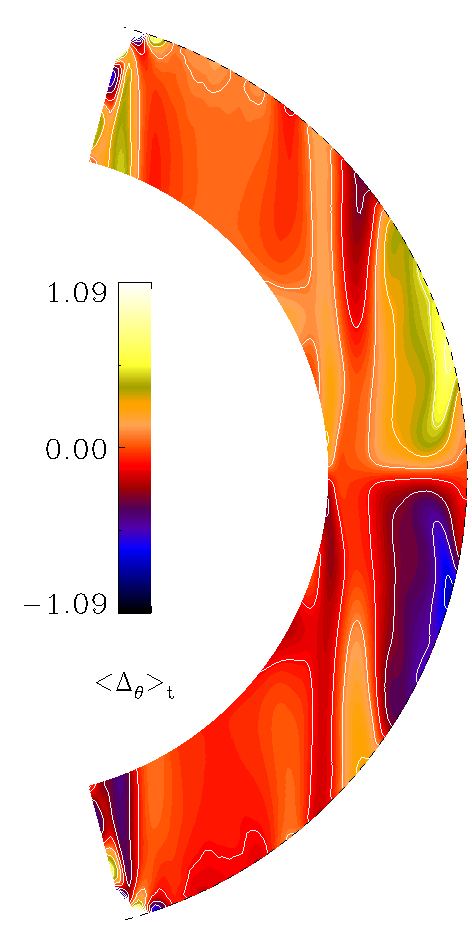}
  \includegraphics[width=0.18\textwidth]{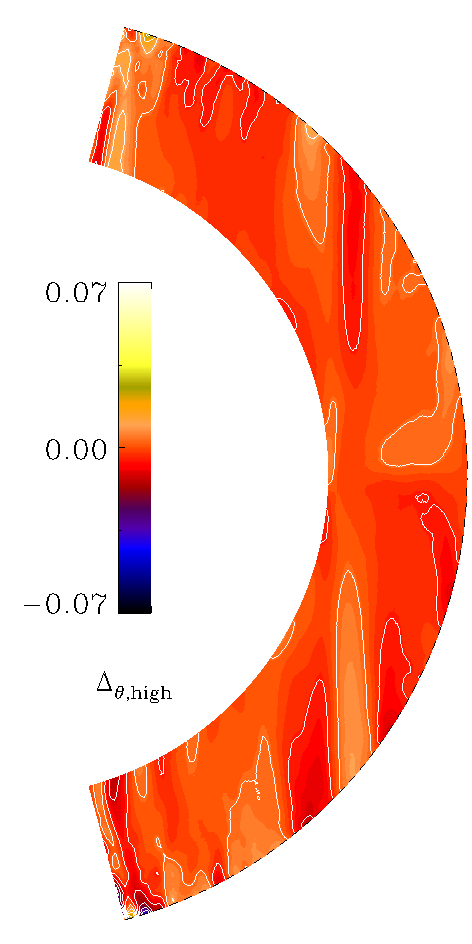}
  \includegraphics[width=0.18\textwidth]{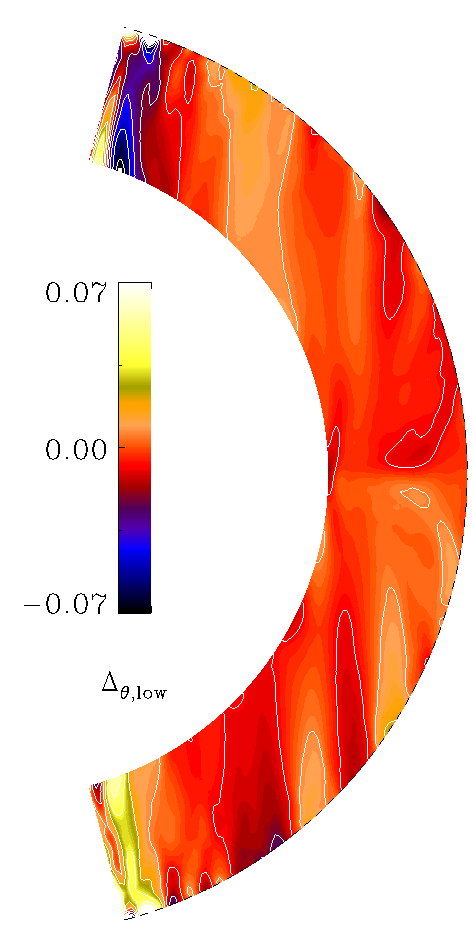}
  \includegraphics[width=0.18\textwidth]{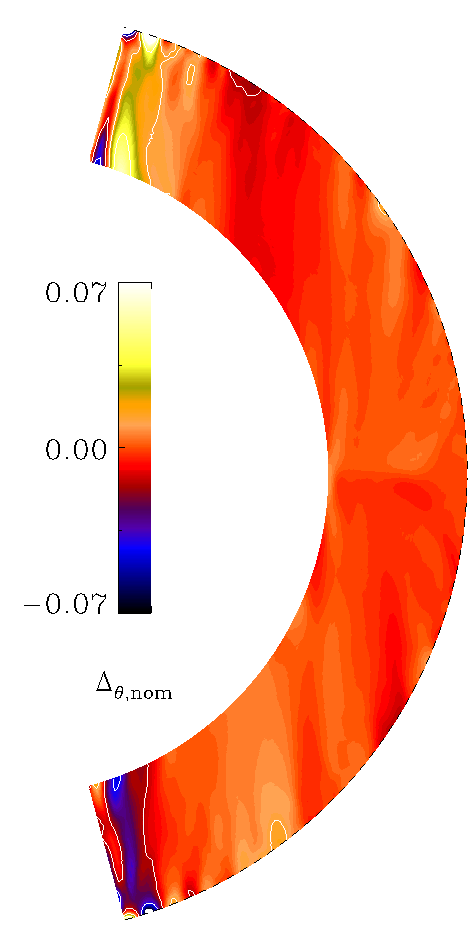}\\
\end{center}\caption[]{Top row, from left to right: normalized
  time-averaged
  rotation profile $\brac{\Omega}_t/\Omega_0$, difference to the
  rotation profile,
  $\Delta \Omega_{\rm state}$, during the
high, low, and nominal state of global activity (D1).  Middle (bottom) row, the
same for mean radial (latitudinal) differential rotation.
}\label{drot}
\end{figure*}

\begin{figure*}[t!]
\begin{center}
  \includegraphics[width=0.24\textwidth]{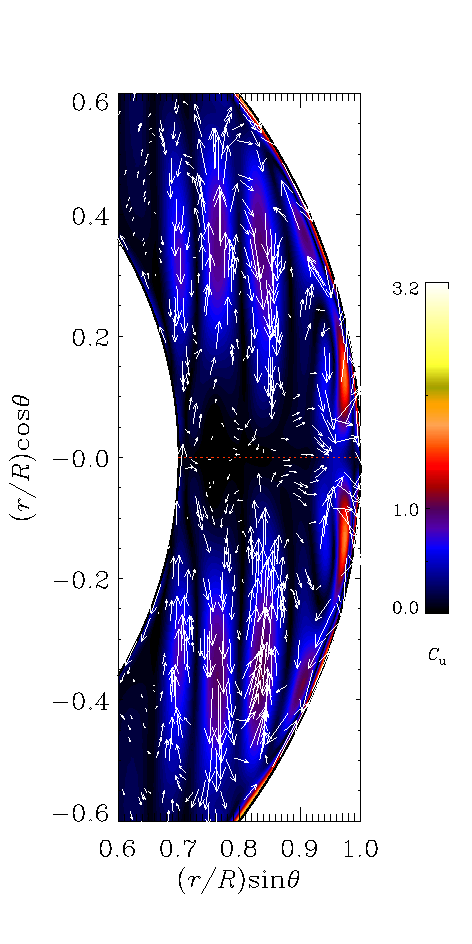}
  \includegraphics[width=0.24\textwidth]{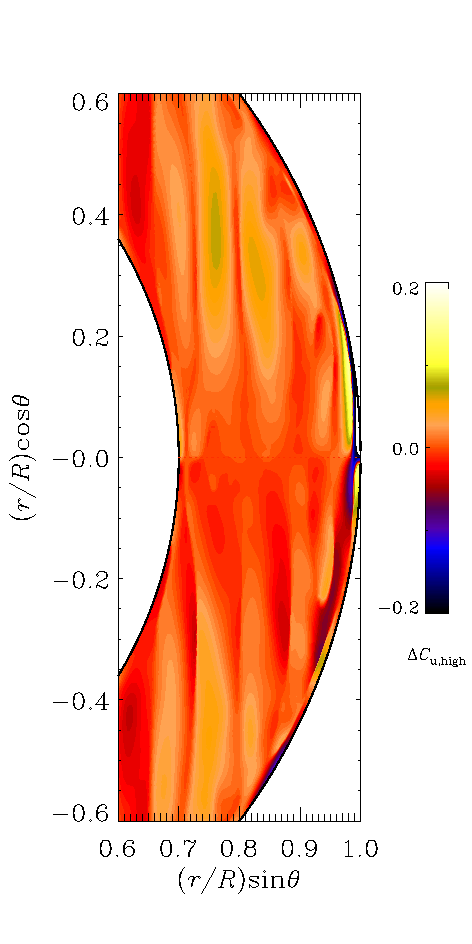}
  \includegraphics[width=0.24\textwidth]{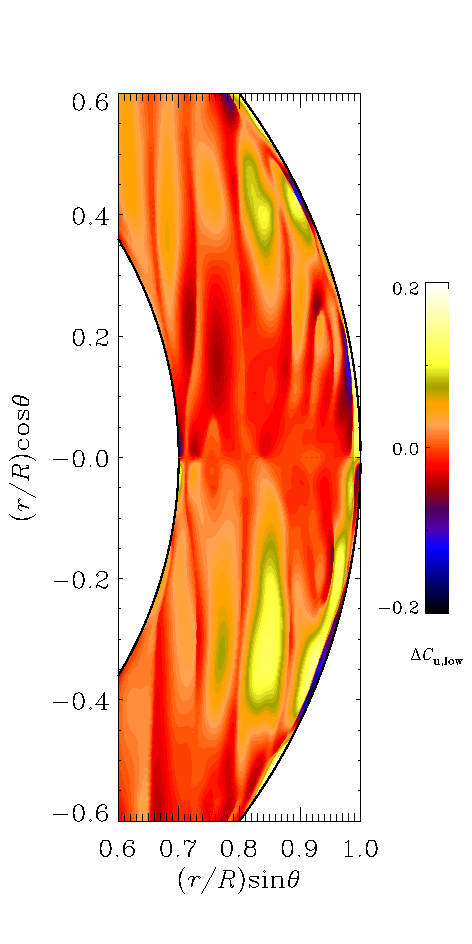}
  \includegraphics[width=0.24\textwidth]{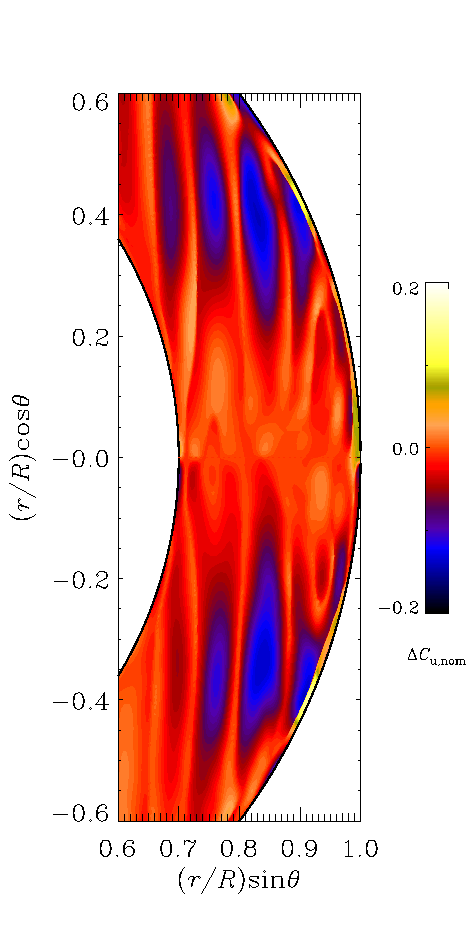}
\end{center}\caption[]{Meridional flow, indicated as arrows, and
  time-averaged dynamo parameter $C_{\rm u}$, indicated with color contours.
On the left, we show the meridional circulation profile averaged over
the
whole simulation time span, and the next three panels show the difference
$\Delta C_{\rm u, \brac{state}}=C_{\rm u}-C_{\rm
  u,\brac{state}}$ to the mean during a high, low, and nominal global activity
state (D1).
}\label{CU}
\end{figure*}

Next we examine the meridional flow profiles and compute a dynamo number
\begin{equation}
  C_{\rm u}=\frac{\mean{u}_{\rm mer}^{\rm rms} \Delta r}{\etatz},
\end{equation}  
where $\mean{u}_{\rm mer}^{\rm rms} = \sqrt{\mean{u_r}^2
  +\mean{u}_\theta^2}$ and other quantities are defined in the same
way as in Eq.~(\ref{Com}). The largest deviation from the
time-averaged profile is, similar to the differential rotation,
obtained when the activity states are defined based on the global
magnetic field energy (D1), the magnitude of the change being
maximally 6 percent in a thin layer concentrated in the equatorial
region near the surface. The results are presented in Fig.~\ref{CU},
zoomed into the equatorial region.  There are, however, some enhanced
meridional flows generated also near the latitudinal boundaries, but
as these are most likely artefacts arising from the latitudinal
boundaries, they are not shown here. The meridional flow pattern
(Fig.~\ref{CU} leftmost panel) is very similar to those obtained
earlier by \cite{KMB12,KMCWB13} and \cite{WKMB13,WKKB15},
i.e.\ consisting of three cells in radius, aligned with the inner
tangent cylinder. Additionally, there is one anti-clockwise cell
confined near the surface and equator, the near-surface poleward flow
being the strongest at this location. The strongest variations in
magnitude over time are related to this region. The dynamo number is
varying spatially between 0...3 averaged over the whole simulation
time span, the dynamo number of the meridional flow being more than
thirty times weaker than of the $\Omega$ effect (radial differential
rotation). The high (second panel in Fig.~\ref{CU}) and low (third
panel in Fig.~\ref{CU}) states show no marked difference in the
spatial distribution nor strength of the main cells, the small
near-equator cell undergoing the strongest variations.  The most
interesting observation is the marked hemispheric asymmetry pronounced
especially during the low state: the meridional flow in the southern
hemisphere is more strongly quenched than the northern one. Similar,
but the effect being more localized to the surface regions, is seen
during the high state, when the northern surface flow gets reduced,
while the southern one gets stronger.

In conclusion, if the MM was an epoch of a global magnetic energy minimum,
our results would predict faster surface rotation, stronger
differential rotation, and
hemispherically asymmetric meridional flow pattern. If actually a
global magnetic energy maximum was occurring in the deeper layers of
the convection zone during the MM, then our prediction would be
reversed to slower
surface rotation and weaker differential rotation, while the
meridional flow would retain its asymmetric character. Neither of
these scenarios is consistent with the actual observations of sunspot
proper motion during the MM \citep{EGT76,RNR93}. Overall, however, our
analysis suggests that the surface activity disturbances have no
direct and/or straightforward relation to changes in the rotational
speed nor the strength of differential rotation.  Moreover, the
meridional circulation is so weak that it is most likely incapable of
influencing the global dynamics significantly.  During the first 100
years of the simulation there are strong dips in the angular velocity,
accompanied by simultaneous maxima in the global magnetic field
strength. Only one of these events (during $t=20$--$45\yrs$), however,
leads to the disappearance of the surface activity. Even then, the dip
in the angular velocity does not coincide with the beginning and the
duration of the disturbed surface activity period. Therefore we can
conclude that the surface irregularities do not originate from the
variations seen in the angular velocity or meridional circulation, at
the same time noting that a disturbed surface activity may not be
indicative of a decrease in the global magnetic energy level.

\subsubsection{Proxy of the $\alpha$ effect}\label{sect:alp}

We build a proxy for the isotropic $\alpha$ effect following the
definition of \citet{PFL76},
\begin{equation}
\alpha = -\onethird \tau (\mean{\mbox{\boldmath
    $\omega^\prime$}\cdot{\bm u^\prime}}
-\mean{{\bm J}^\prime\cdot{\bm B}^\prime}/\mean{\rho}),\label{eq:Calp}
\end{equation}
where the current helicity consists of
the fluctuating current ${\bm J}^\prime=\nab\times{\bm B}^\prime/\mu_0$ and
the fluctuating magnetic field ${\bm B}^\prime={\bm B}-\mean{\bm B}$.
When looking at the kinetic and magnetic contributions to the $\alpha$
effect proxy, it can be observed that the kinetic helicity is
dominating the signal, the current helicity being roughly an order of
magnitude weaker. Therefore, we mostly concentrate in the following on
examining the properties of the kinetic helicity, while also computing a
suitable dynamo number.

First we examine the time evolution of the azimuthally averaged
kinetic helicity $H_{\rm kin}=\mean{\mbox{\boldmath $\omega^\prime$}\cdot{\bm u^\prime}}$.
This quantity shows a clear mean signal as a function of latitude, so
that negative (positive) values are obtained in the north (south),
increasing towards the poles where maximal values are obtained, with a
strong localized enhancement seen in both hemispheres in the low-latitude
regions, extending at least half of the depth of the convection
zone. Close to the bottom of the convection zone at high-latitude
regions, the sign of this quantity is reversed. These results are in
agreement with earlier ones both in local and global domains
\citep[e.g.][]{KKB09a,KMCWB13}. In Fig.~\ref{HelZoom}(a), we
show this quantity with the mean
profile over the whole time series subtracted, showing
that most of the variation occurs near the equatorial
region. The signal is dominated by the high-frequency modes (M1 and
M2), while at higher latitudes the dominant mode M7 in the magnetic
field is detectable, but not significant in comparison to the
high-frequency signal.
The zoomed-in plots of Fig.~\ref{HelZoom}(b) and (c) reveal the presence
of the weak basic cycle modulation in the kinetic helicity, but with
half of the length of the magnetic cycle. Especially
during times of clear antisymmetry, Fig.~\ref{HelZoom}(d), a
modulation pattern is seen near the surface both at low latitudes
(location of the strongest signal in kinetic helicity, shown with
solid lines in the figure), and at higher latitudes (location of the
toroidal magnetic field maximum; dashed lines). The modulation pattern
is such that the extrema of helicity roughly coincide with those of
the radial magnetic field at mid-latitudes (shown with dotted lines in
the figure). During times when M7 shows stronger symmetry with respect
to the equator, Fig.~\ref{HelZoom}(e), all the
observed weak dependencies in the earlier antisymmetric phase of the
simulation break down, although signs of immediate restoration of the
modulation pattern can be detected when the symmetry abruptly switches
at around $t=234\yrs$ back to antisymmetric one. Interestingly, signs of a
longer-term modulation are seen in Fig.~\ref{HelZoom}(d), as during
that epoch the kinetic helicity signal is systematically more negative
(positive) in the north (south) in comparison to the mean value over
the whole time span. This is also manifested in Fig.~\ref{HelZoom}(b)
as persistent blue (yellow) stripes at low northern (southern)
latitudes.  Detected by EEMD only through the high value of the
regularity measure, however, this modulation is of very low energy in
comparison to the high-frequency modes.

\begin{figure*}[t!]
  \begin{center}
  \includegraphics[width=\textwidth]{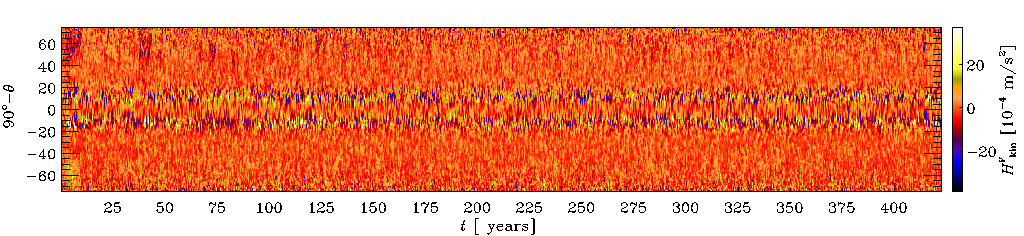}\\
  \includegraphics[width=\textwidth]{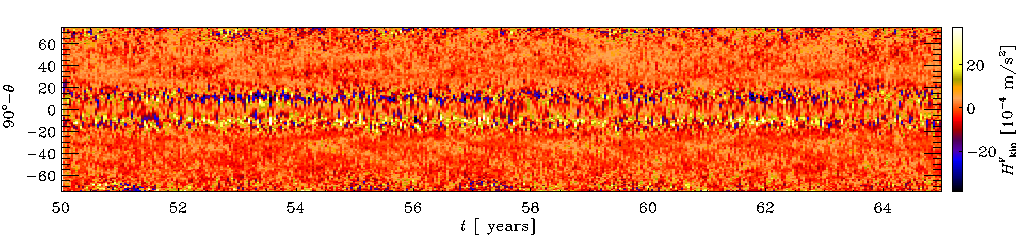} \\
  \includegraphics[width=\textwidth]{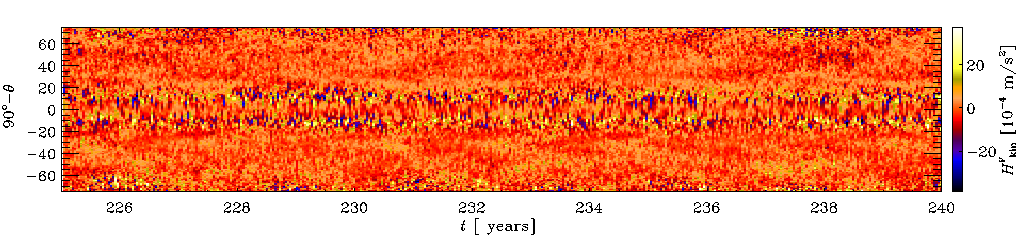}\\
    \includegraphics[width=0.46\textwidth]{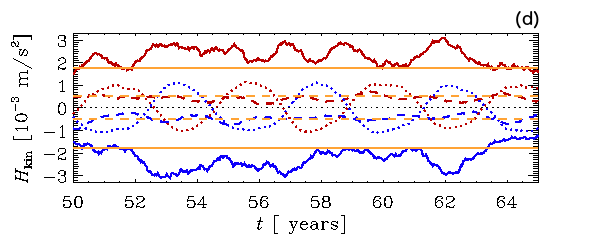} \includegraphics[width=0.46\textwidth]{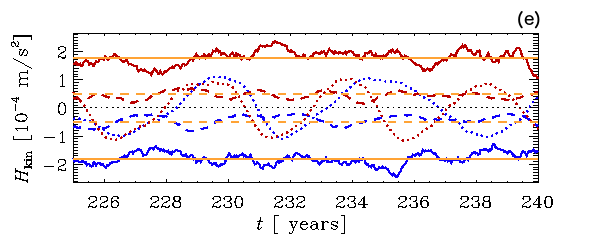}\\
 \end{center}\caption[]{
(a) Time-latitude plot of the kinetic helicity variations $H_{\rm
  kin}^{\rm v}=H_{\rm kin}-\brac{H_{\rm kin}}_t$.
    Zoomed-in plots of $H_{\rm kin}^{\rm v}$
over $t=50$--$65\yrs$ (b) and $t=225$--$240\yrs$ (c).
(d) $H_{\rm kin}$ at
$\pm 10^\circ$ (solid) and $\pm 25^\circ$ (dashed) latitude for north (blue) and south
(red) with the means over the whole time series indicated with an
orange horizontal lines. The data is averaged with a running mean over
a 1-year window, to smooth out the dominant high-frequency
component. The dotted lines show the radial magnetic field at $\pm
25^\circ$ latitude, scaled to fit the plot.
All quantities are plotted at $\Rs$.
  }\label{HelZoom}
\end{figure*}  

\begin{figure*}[t!]
\begin{center}
  \includegraphics[width=0.24\textwidth]{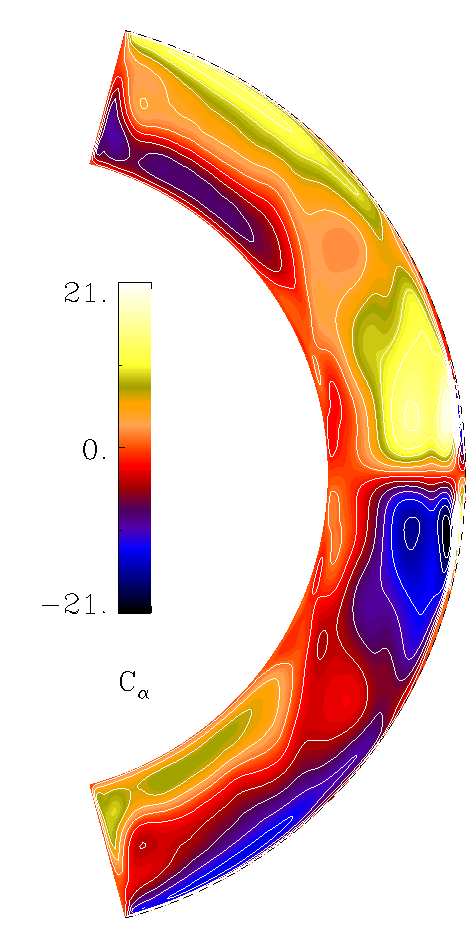}
  \includegraphics[width=0.24\textwidth]{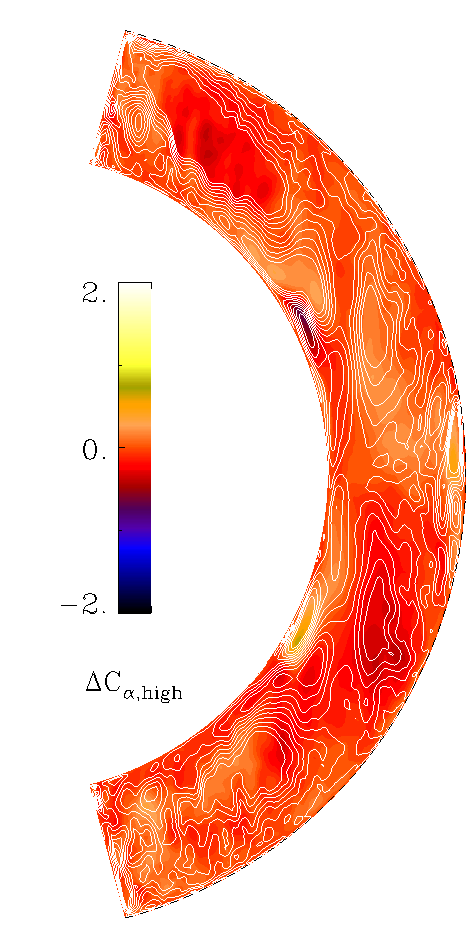}
  \includegraphics[width=0.24\textwidth]{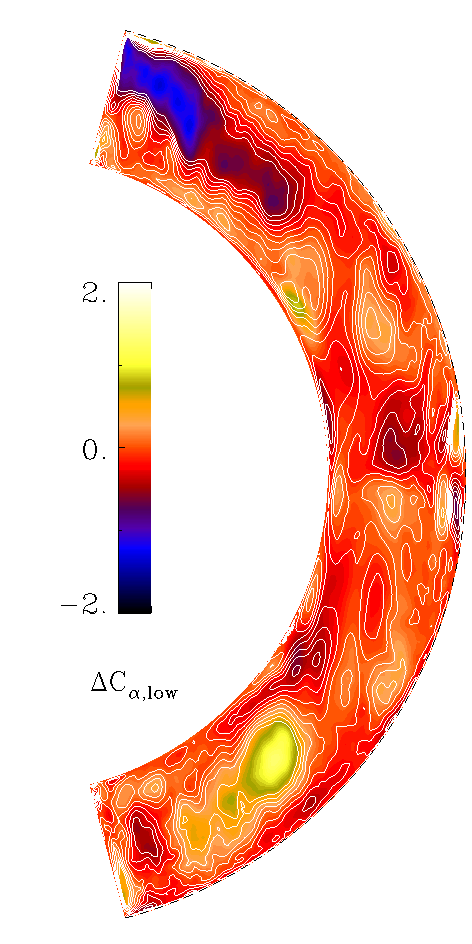}
  \includegraphics[width=0.24\textwidth]{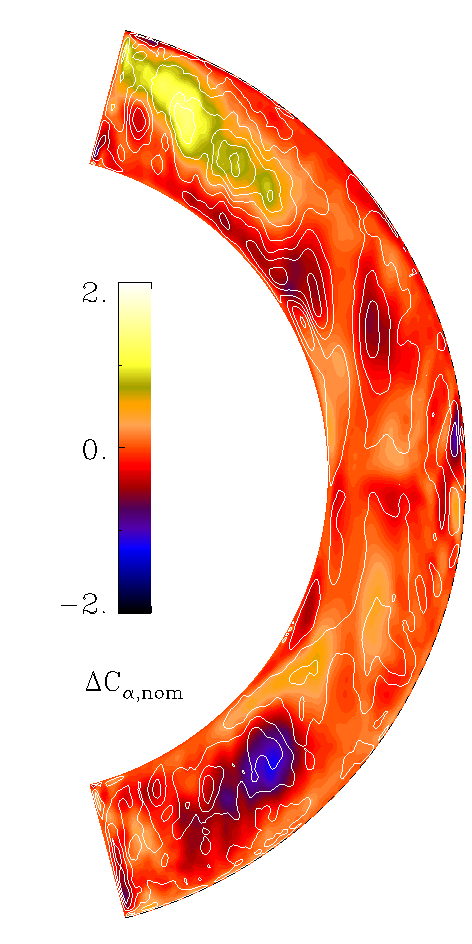}
\end{center}\caption[]{
  From left to right: $C_\alpha$ time-averaged over saturated stage.
  Difference to the $C_\alpha$ profile during
  the high, low, and nominal state of surface magnetic activity (D2).\label{Calp}}
\end{figure*}

Next, we define and compute a dynamo number
\begin{equation}
  C_\alpha=\frac{\alpha \Delta r}{\etatz}.\label{eq:Calp}
\end{equation}  
This quantity is plotted in
Fig.~\ref{Calp}, left panel, as the time average over the whole length
of the simulation. The profile is similar to those obtained earlier,
e.g.\ in Runs C1 and C2 of \citet{KMCWB13}.

This dynamo number is almost an order of magnitude larger than the
corresponding one from meridional circulation, but roughly 5 times
weaker than the one estimated from radial differential rotation.
The strongest deviation from
the mean profile is obtained when high and low states are determined
according to the surface activity level (D2), while selecting the
time points based on the global (D1) and bottom energy (D3)
levels produce no significant deviation from the mean.
During the low state, the $\alpha$ effect is strongly enhanced
i.e.\ significantly larger positive/negative values are obtained in
these regions in north and south, respectively.
During the high state, there is also a mild enhancement, while in the
nominal state the $\alpha$ effect is strongly reduced. The magnitude
of the enhancement/quenching is the largest of all the dynamo drivers,
roughly by 30 percent, but acts to the opposite direction as expected,
i.e. with a locally and temporally larger $C_\alpha$, the naive
expectation is to obtain a more efficient dynamo near the surface, but
weaker surface activity is seen.

We also compute the migration direction of the magnetic field
predicted by the Parker-Yoshimura sign rule \citep{Yo76}
\begin{equation}
  \bm{\xi}_{\rm migr}=-\alpha \hat{\bm{e}}_{\phi} \times \nabla \bm{\Omega} ,\label{eq:eqmigr}.
\end{equation}
We plot this quantity on top of the rms-value of the mean toroidal
magnetic field in the convection zone in Fig.~\ref{ReM}, top panel.
In the entire region of negative radial shear, manifested by the
region of slower rotation in the differential rotation profile,
equatorward migration is predicted to occur in both hemispheres. This
region coincides with a toroidal field belt having a maximum roughly
at $\pm 28^\circ$ latitude and $0.85\,R_\odot$ depth. There is another
equatorward migration region at the very bottom of the convection
zone, coinciding with the location of the strongest toroidal field
belt roughly in a similar latitude range. In the upper third of the
convection zone, however, the predicted migration direction is
poleward. This matches with a toroidal magnetic field belt at $\pm
22^\circ$ latitude and $0.9\,R_\odot,$ depth.  No strong radial
migration is predicted for the lower parts of the convection zone,
within which both upward and downward regions are seen, but without a
clear systematic pattern.  The good agreement of the predicted and
actual migration direction as well the migration pattern itself fit
well with the work of \cite{WKKB14,WKKB15}.  The migration pattern
does not change significantly using any of the activity level
definitions (D1--D3), which is likely due to the fact that the
gradients of the angular velocity and the kinetic helicity, and
therefore the $\alpha$ effect, are sensitive to different activity
indicators (the former to the global, the latter to the surface) and
the signal is cancelled out.  Therefore, we note that the $\alpha$
effect shows the most interesting behavior during the extrema seen in
the surface activity, but our approach to investigate its effects via
simple scalar proxies of kinetic and magnetic helicities is
inadequate; we will return to this problem in Paper II.

Finally, to investigate whether the different lengths of the dynamo solutions
and their distinct spatial distribution is due to a systematic, strong
dependence of the magnetic diffusivity or magnetic Reynolds number
\begin{equation}
\Rem(r,\theta)=\frac{\alpha_{\rm MLT} H_p \urmsp(r,\theta)}{\eta},
\end{equation}
where $\eta=10^{8}$~m$^2$~s$^{-1}$ is the molecular diffusivity, 
  profile either on radius or latitude, we plot their radial profiles at
two different latitudes in Fig.~\ref{ReM}, lower panel. To explain the
co-existence of dynamo solutions with more than two orders of
magnitude varying periods between the bottom and the top with the
dependence being due to a weaker diffusivity in the bottom parts, a
two orders of magnitude increase of the magnetic Reynolds number as
function of radius should be seen. Even though the trend seen is
indeed increasing as function of depth, the magnitude of the increase
is maximally four, being too weak to provide an explanation.

\begin{figure}[t!]
  \begin{center}
  \includegraphics[width=0.7\columnwidth]{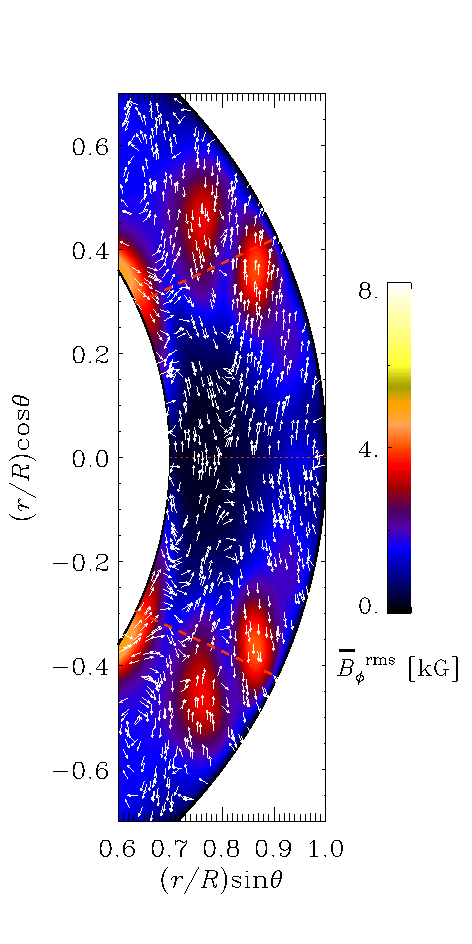}\\
  \includegraphics[width=\columnwidth]{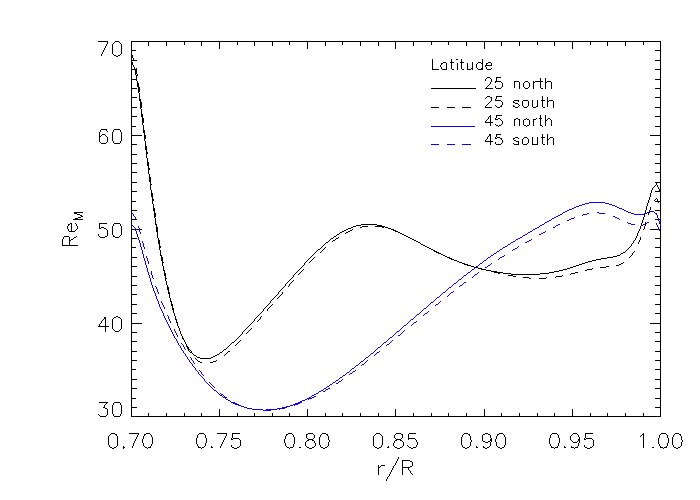}
  \end{center}\caption[]{
  Top panel: color contours of $\meanv{B}^{\rm rms}_{\phi}$ in kGauss
  averaged over the whole simulation run, zoomed-in to show the near
  equatorial region. The overplotted arrows show the predicted (Eq.~\ref{eq:eqmigr}) time-averaged
  mean migration
  direction ${\bm\xi}_{\rm migr}$ normalized to unity.
  The dashed red lines show the radius at $\pm25$ degrees of latitude. 
    Lower panel: radial dependence of the magnetic Reynolds number at two latitudes (north -- solid; south -- dashed.}
\label{ReM}
\end{figure}

\subsection{Equatorial symmetry}\label{sect:NS}

In Fig.~\ref{parity} we plot the equatorial symmetry of the magnetic
field, defined as
\begin{equation}\label{eq:eqsym}
P=\frac{E_{\rm even}-E_{\rm odd}}{E_{\rm even}+E_{\rm odd}},
\end{equation}
where $E_{\rm even}$ is the energy of the quadrupolar (symmetric) and
$E_{\rm odd}$ the energy of the dipolar (antisymmetric) mode of the
magnetic field. In Fig.~\ref{parity}(a) we show the
global parity as a function of time. This quantity is obtained by
computing, for each latitude pair for certain depth, the energies
contained in the even and odd modes, and taking the mean of the
obtained data. Evidently, the parity is
mixed throughout the whole simulation, the dominant mode being the
dipolar (solar-like) mode (average parity $-0.15$ over the whole time
series). A difference can again be observed between the first 3/4 of
the simulation (average parity of $-0.26$) in comparison to the last
quarter, when there are larger fluctuations around
the clearly positive mean parity of $0.1$.

The best correlation between the different activity level definitions
can be obtained when D3 is used. The parity not only shows a clear
modulation with a similar long-term cycle as the toroidal field in the
bottom of the convection zone, but also a clear pattern is revealed
for the first 3/4 of the simulation with the D3 color coding: during a
bottom mode minimum, the parity also obtains a minimum ($-0.42$), while
during a high state, the parity is the least solar-like
($-0.14$). During the last quarter of the simulation, the bottom mode is
so weak, that it falls below the low activity limit for the whole
epoch. Nevertheless, the long-term cycle persists both in the bottom
mode and in the parity, suggesting a relation between these
quantities.

As can be seen from Table~\ref{mode_summary}, the low-frequency bottom mode (M11)
is predominantly of dipolar symmetry (mean parity $-0.36$).
From Fig.~\ref{parity}(b), it is evident that rather strong
parity fluctuations are related to this mode, the excursions to positive
parity being more short lived than the periods spent in the negative
parity region.
The majority
of the parity variation, however, is related to the dominant `basic'
cycle (M7), which on average has a parity of roughly $-0.12$,
but its symmetry is strongly fluctuating as a function of time, as is
evident from Fig.~\ref{parity}(b).
The parity is fluctuating even more strongly than that of M11, between
nearly purely dipolar and quadrupolar states. Most of the time M7 and
M11 and in anti-phase, i.e. when M11 attains the largest positive
value, M7 is the most negative. There is a tendency for irregular
behavior when the parities of these modes are in phase, such as during
$t=20$--$45\yrs$ and $t=250$--$300\yrs$.
In Fig.~\ref{parity}(c) and (d) we show two zoom-ins of the parity
evolution, panel (c) showing a clearly regular, antisymmetric behavior
of the system and the surface toroidal field ($t=50--65\yrs$), while
panel (d) shows a predominantly symmetric state that during only a
half a cycle changes into a antisymmetric one
($t=225$--$235\yrs$). During the former epoch, the bottom toroidal magnetic
field strength attains a minimum, while during the latter, a maximum
is observed.}

\begin{figure}[t!]
\begin{center}
\includegraphics[width=\columnwidth]{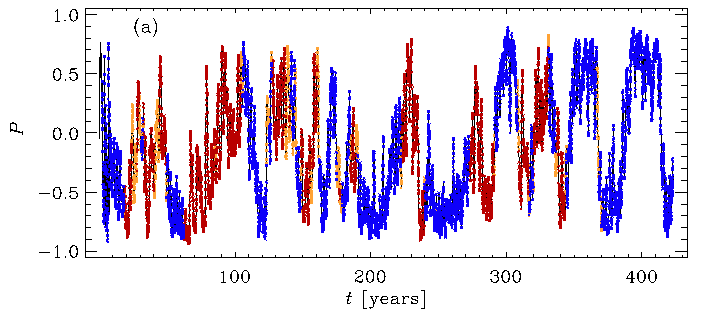}\\
\includegraphics[width=\columnwidth]{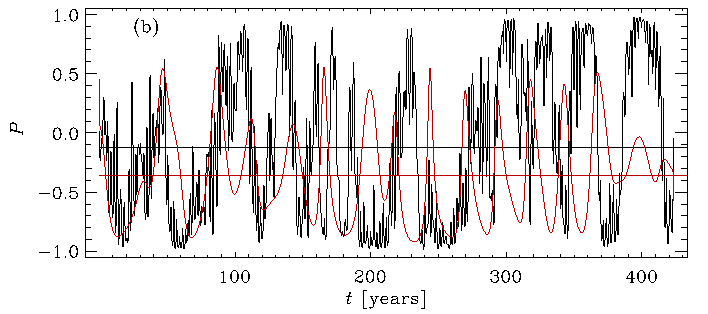}\\
\includegraphics[width=\columnwidth]{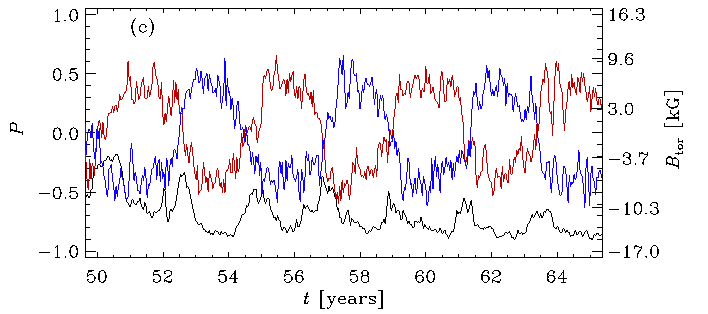}\\
\includegraphics[width=\columnwidth]{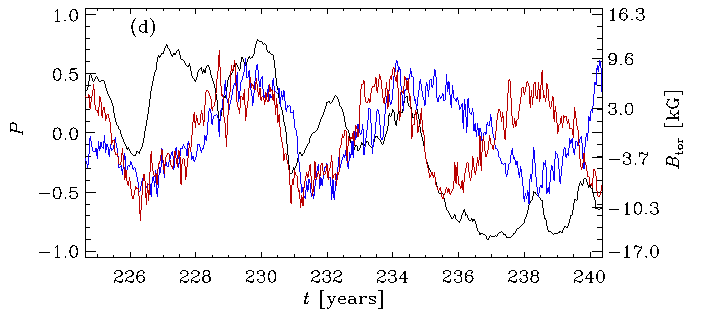}\\
\end{center}\caption[]{(a) Parity $P$ as function of time. The
    color coding is defined using Definition D3.
  (b) Parity of M7 (black) and M11 (red) as function of time;
  their means are indicated with the horizontal lines with the same color coding.
  (c) Parity (black) and toroidal magnetic field near the surface (blue: north,
  red: south) at $\pm25^\circ$ latitude, during the low state of the
  bottom toroidal mode, panel (d) shows the same quantity, but for
  a high state of
the bottom toroidal mode.}
\label{parity}
\end{figure}

\begin{figure}[t!]
\begin{center}
  \includegraphics[width=\columnwidth]{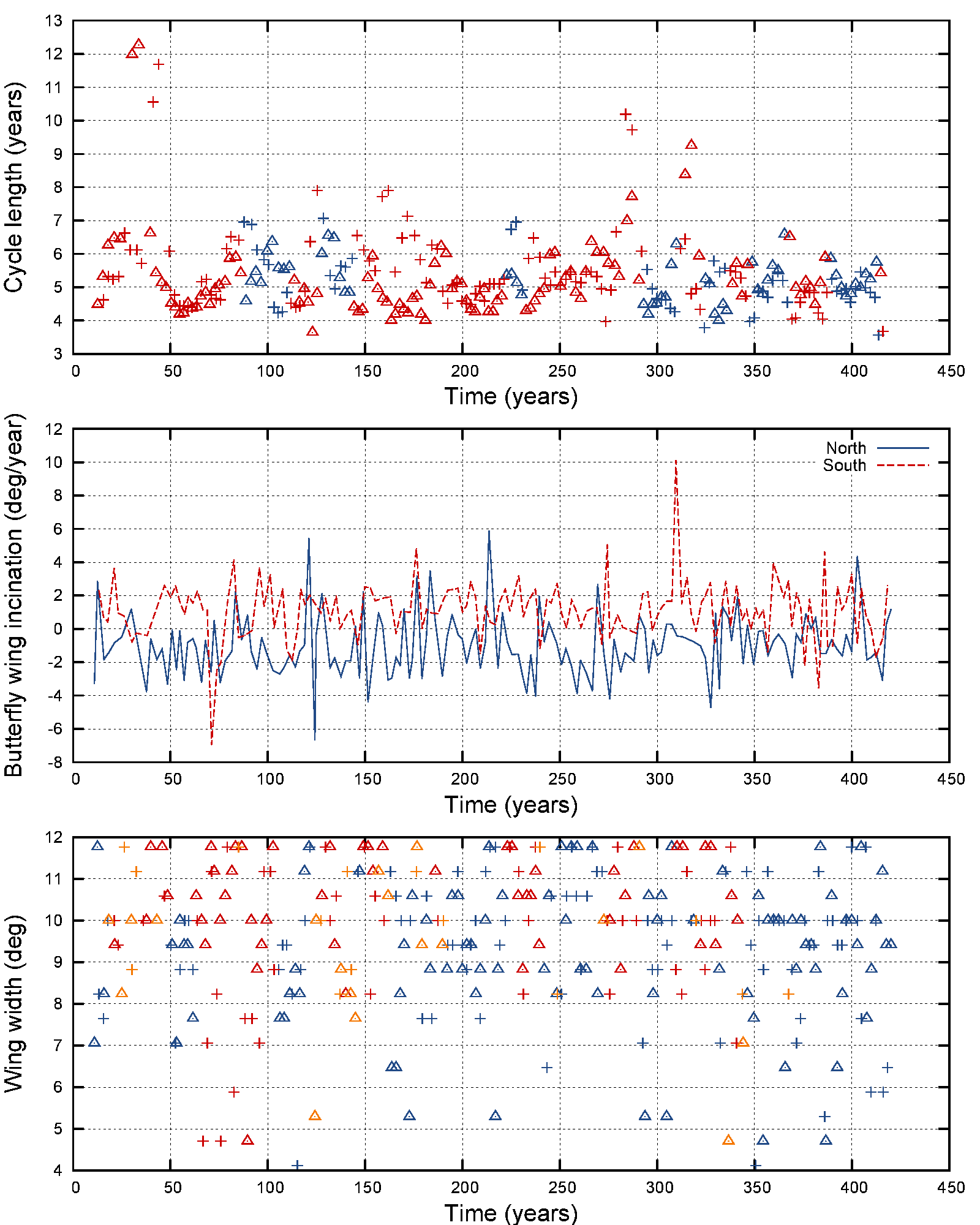}\\
\end{center}\caption[]{
  Top panel: magnetic cycle
  length determined from the toroidal magnetic field, $\mean{B}_\phi$,
  at $\pm25^\circ$ latitude and $r=\Rs$, blue/red indicating
  positive/negative parity,
  north with triangle and south with plus symbols. Middle panel: time evolution of butterfly wing inclination.
  Bottom panel: butterfly wing widths, red/orange/blue indicating high/nominal/low bottom magnetic field energy (according to D3),
  north with triangle and south with plus symbols.} \label{cycleprop}
\end{figure}

\subsection{Variation of general cycle properties}\label{sect:cycle_prop}

Evidently, the overall properties of the magnetic cycles show quite a
large variation over time. In Fig.~\ref{cycleprop} we plot three
indicators, namely the cycle length of the toroidal field
near the surface at $\pm25^\circ$ latitude, the inclination of the
butterfly wing, and the butterfly width in degrees for north (triangle
symbols) and south (pluses) separately.

As for the toroidal field cycle length (Fig.~\ref{cycleprop}, top
panel), two distinct types of events can be seen: During
suppressed magnetic activity near the surface layers ($t=$20--45,
275--285, and 310--320$\yrs$), the field ceases to change its sign;
therefore, some cycles are `missed', and then twice as long cycles are
detected as a result. The other type of variation are the less abrupt
fluctuations around the mean cycle length, the shortest cycle lengths
being slightly less than 4 years and the longest one around around 8
years. The cycle length is not sensitive to any of the activity level
definitions (D1--D3), but there is a tendency for the cycles to be
shorter if the parity is more symmetric (blue symbols), while larger
cycle length variations occur for more solar-like (antisymmetric; red
symbols) parities. Also, all the `missing' cycles occur during the more
solar-like parity, during which the bottom cycle is attaining its
minimum.

To further analyze the general cycle properties we have filtered the
surface magnetic field data so that only points with
$|\mean{B}_{\phi}| > 4kG$ have been retained.  This way the
concentrations of strong magnetic field are clearly grouped into
separate half-cycles or wings. The relative properties of these
structures, such as the latitudinal extent (width) and duration can
then be directly measured. Based on these properties we define the
inclination of the butterfly wing as the half cycle width divided by
its duration.  For north/south, a solar-like cycle would have
negative/positive inclinations, while for cycles where migration is
weak or non-existent, values close to zero are expected. As is evident
from the middle panel of Fig.~\ref{cycleprop}, there is a tendency for
the `missing' cycles to appear with weak migration only. Anti-solar
cycles are also observed quite a number of times both for north and
south, but not at simultaneous epochs. It appears that the shorter the
cycle, the higher is the probability for anti-solar migration pattern.

The corresponding butterfly width (Fig.~\ref{cycleprop} bottom panel)
is most often large for the `missing' cycles, which is in striking
contrast to the observational evidence for the MM \citep[see][and
  references therein]{Usoskinetal15}, if these two types of events are
in reality representing one and the same phenomenon. There is
a tendency for narrow widths to occur when the bottom magnetic field
is weak.

\begin{figure*}[t!]
\begin{center}
  \includegraphics[width=\textwidth]{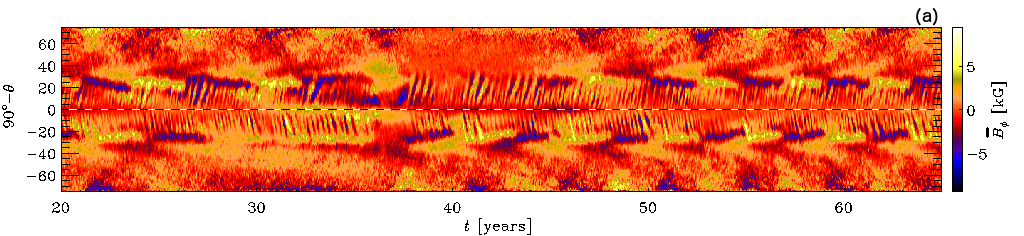}\\
  \includegraphics[width=\textwidth]{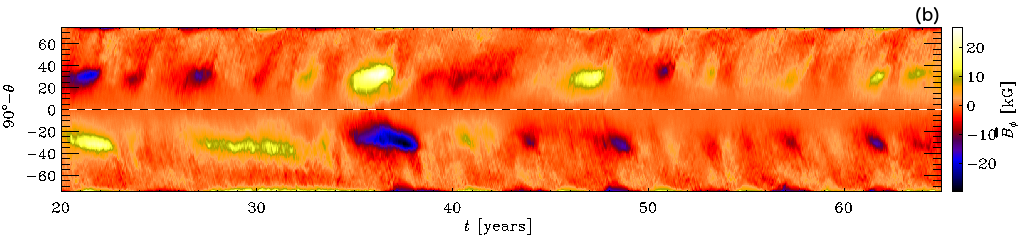}\\
  \includegraphics[width=\textwidth]{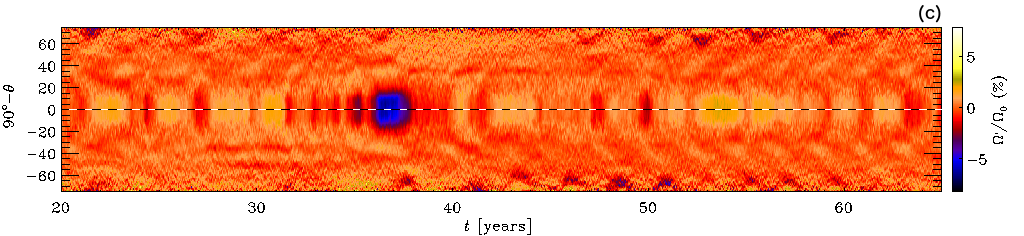}\\
  \includegraphics[width=\textwidth]{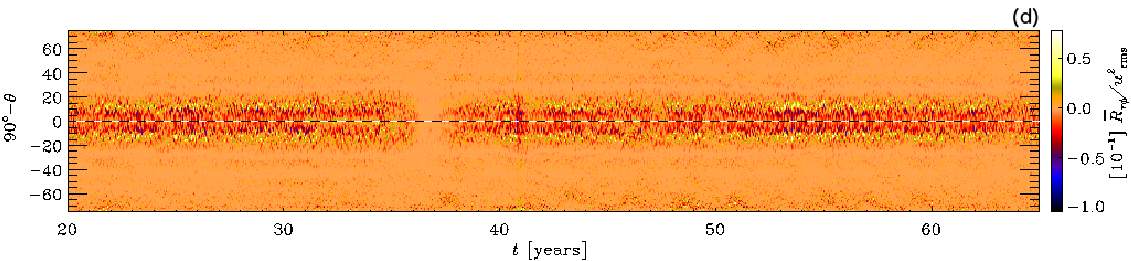}\\
  \includegraphics[width=\textwidth]{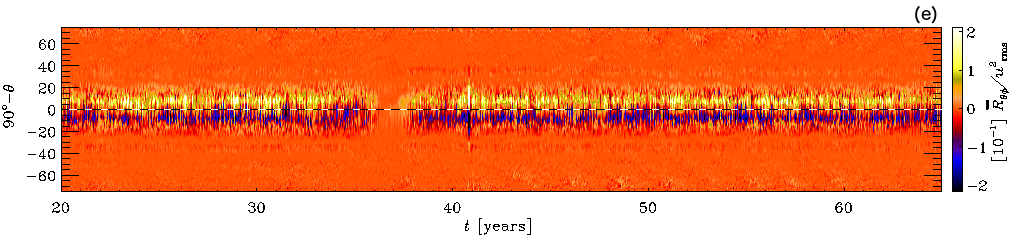}  
\end{center}\caption[]{From top to bottom: Zoom-ins of the time-latitude
diagram of the mean toroidal magnetic field near the surface (at $\Rs$)
and bottom (at $\Rb$),
the angular velocity variations, and the Reynolds stress components
$R_{r \phi}$ and $R_{\theta \phi}$ near the surface (at $\Rs$) during
$t=20$--$65\yrs$, showing that the magnetic cycle and surface activity
are strongly disturbed ($t=20$--$45\yrs$), after which the regular state
is quickly recovered ($t=45$--$65\yrs$).}
\label{zoomin}
\end{figure*}

\subsection{Irregular event during 20--45 years}\label{sect:MM}

Finally, we present a close-up of the most clearly pronounced
irregular activity epoch during 20--45 years. In
Fig.~\ref{zoomin}, we reproduce the plot shown in
Fig.~\ref{torsional}, but restrict the plotting range to 20--65 years,
which shows both the irregular evolution, followed by the resumption of the regular
behavior, indicating that the system changes itself into very abruptly after the
distortion. The time resolution in the figure is high enough to
clearly show the high-frequency (M1) mode of the magnetic field (top),
not well discernible in any other longer time span figure; this
cyclicity is seen to
persist regardless of the irregularities seen in the modes M7 and M11.

The irregular behavior is first seen in the southern hemisphere, see
Fig.~\ref{zoomin}(a),
and the event appears to be
preceded by a field enhancement in the bottom mode (M11;
Fig.~\ref{zoomin}(b)).
Even though the toroidal field of the mode M7 gets very weak,
the surface activity does not totally cease, but the field evolves
with significantly longer period without a polarity change. The
southern cycle, therefore, appears truly as a `missed' one; meanwhile
the northern hemisphere continues as usual. After twice
the time of a normal cycle has elapsed, the bottom magnetic field
attains another strong maximum. This is reflected by a strong minimum
in the Reynolds stresses, Fig.~\ref{zoomin}(d) and (e), and surface
differential rotation, Fig.~\ref{zoomin}(c). Immediately after this,
the toroidal magnetic field practically vanishes from the northern
hemisphere, while the south seems totally unaffected by the northern
disruption. Again, after the south has produced two normal-looking
magnetic cycles, normal activity is rapidly restored also in the
north. After that, the bottom magnetic field exhibits no further
strong extrema, and the mode M7 continues its evolution in a regular
manner. It is also noteworthy that the bottom magnetic field changes
its polarity during the disturbed epoch.

During this epoch, it is evident even by eye that there is especially
strong equatorial (north-south) asymmetry related to the surface
toroidal magnetic field. This is reminiscent of the observations of
the Maunder and Dalton minima \citep{RNR93,UMK09}, that indicate
strong hemispheric asymmetry and a relatively strong quadrupolar
component of the magnetic field.  The global parity does not strongly
reflect this local surface anomaly as a sudden, persistent change
towards a quadrupolar state (+1), although the quantity fluctuates
strongly between the values [-0.9,0.7]. This is most likely due to the
fact that the symmetry properties of the very strong bottom toroidal
magnetic field dominate the signal, at all times remaining close to a
dipolar configuration.

Although similarly dramatic events cannot be isolated from the
rest of the time series, it is clear that both extrema and polarity
reversals of the bottom mode M11, frequent during the first 3/4 of the
simulation time, make the system more irregular than during the last
1/4 during which especially the energy contained in the bottom mode
M11 gets weaker.

\section{Conclusions}

In this paper we have analyzed a semi-global (wedge-shaped) DNS that
produces a solar-like oscillatory dynamo with surface migration
properties of the magnetic field resembling those from
observations. The mean cycle length is 4.9 years, and the simulation
is evolved over 80 such magnetic cycles. If scaled to solar units
using the fact that the Sun has a magnetic cycle of roughly 22 years,
our simulation would correspond to two millennia of solar evolution.
The chosen parameters, most notably the modest stratification and the
values of the Reynolds numbers, are still far from the real Sun;
nevertheless, this combination of parameters produces a solar-like
dynamo, and it is computationally affordable to integrate over a large
number of cycles.  The rotation rate is five times solar, resulting in
a roughly three times weaker differential rotation than in the
Sun. The rotation profile is solar-like, but somewhat more
cylindrical, and the meridional circulation consists of multiple cells
in radius. Nevertheless, we can expect that this simulation can well
be used to study the long-term evolution of solar-type dynamos.

A general property of the dynamo solution is its cyclic
nature: even though there is a clear magnetic cycle, the various
nonlinearities in the system drive it away from an exactly periodic
(harmonic) state. Therefore, special time series analysis techniques
are needed that can cope with non-periodic signals. In this study, we
perform a statistical analysis over the whole convection zone,
investigating the properties of all relevant quantities at different
depths and latitudes. The methods we choose to employ are the
EEMD and $D^2$ statistics.

The behavior of the dynamo solution is extremely complex. In addition
to changing cycle length, we observe epochs of disturbed and even
ceased surface activity, and strong short-term hemispherical asymmetries.
The major general findings related to the overall dynamics of the system include:
\begin{itemize}
\item The hemispheric asymmetries are related to the
  magnetic field
  alone, while the velocity field remains almost perfectly
  symmetric at all times.
\item The epochs when the surface activity has decreased or is practically
  non-existent are not global magnetic energy minima but maxima, as
  strong magnetic fields are stored in the deeper parts of the
  convection zone. 
\end{itemize}

The main goal of this study is to find the causes for the irregular
behavior rather than to make direct comparisons with the observed
characteristics of the solar magnetic cycle. Therefore, we have
concentrated our efforts toward analyzing the dynamo solution itself to
investigate how the most important properties (cycle length, migration
properties, energetics) change during the irregular epochs. We also
investigated all the key factors in the dynamo process, namely the
rotation and its non-uniformities, meridional circulation, the
inductive action arising from turbulent convection ($\alpha$ effect),
and the changes of these as functions of a relevant activity level
measures. The major findings from this part of the analysis can be
summarized as follows:
\begin{itemize}
\item The specific dynamo solution analyzed here contains not only
  one, but three separate modes, having distinct cycle lengths and
  symmetry properties, and are located in different parts of the
  convection zone. The dominant mode is the near-surface 4.9 year
  cycle (denoted
  with M7) showing equatorward migration at lower latitudes and
  poleward migration at higher latitudes. This mode is accompanied
  by a weaker high-frequency poleward migrating mode (M1, 0.11 years)
  in the equatorial region and a low-frequency mode at the bottom of
  the convection zone (M11, roughly 50 years).
\item The crucial property of the different dynamo modes is their
  different symmetries. While the dominating surface mode has
  a mixed equatorial symmetry that undergoes strong fluctuations over
  time, the bottom mode exhibits nearly pure antisymmetry and is also
  more regular than the surface mode.
\item There is a close relationship between the global magnetic
  and kinetic energies such that strong magnetic fields quench
  the flow field in a manner that the angular velocity is
    significantly reduced, differential rotation gets somewhat weaker,
    and the meridional circulation attains an asymmetric character
    with modifications of the circulation magnitude especially in
    near-equatorial surface regions. We expect that this is what would
    have been seen during the MM, if it was an event corresponding to the
    abrupt disappearance of surface activity with a simultaneous
    global energy maximum in the bottom of the convection zone
    \cite[see e.g.][for similar arguments]{KAR99,Ka10}.
  \item A more common way of interpreting MM is an overall drop of the
  magnetic activity level. Our simulation produces no such events,
  which is naturally no proof of them not occurring in the real
  Sun. The only quantity markedly linked to the surface magnetic
  activity level in our model is the magnitude of the $\alpha$
  effect.
\item Two kinds of irregularities are separable from the dynamo
  solution itself: smooth variations in the cycle properties and
  abrupt changes that lead to `missing' cycles and ceased surface
  activity. All these can be satisfactorily explained as the interplay
  of the different dynamo modes and their influence on the flow field.
\end{itemize}

Even though this simulation has been run for a considerable length of
time, we still observe a secular trend -- especially for the long-term
cycle in the bottom of the convection zone. To fully capture and
understand this trend, and to find out whether it is a transient or a
real cycle, the simulation would still need to be continued
further. Although we have presented an already rather elaborate and
time consuming data analysis of the results, we have merely touched
upon the turbulent quantities. Especially the $\alpha$ effect was
treated in a very simplified manner, only using a proxy from the
kinetic helicity, which is far from adequate. We recently undertook an
effort to compute the full tensorial representation of both the
diffusive and anti-diffusive contributions to the electromotive force
using the spherical test-field method \citep{WRKKB16} from a rather
similar, but shorter, solar-like dynamo simulation. It would therefore
be useful to compute and compare the turbulent transport coefficients
from the different epochs of the long run presented here.  The
simplified analysis presented here, however, re-affirms our earlier
conclusion of the general dynamo solution and its migration being
explicable in terms of a turbulent $\alpha \Omega$ dynamo
\citep{KMB12,WKKB14}.

\begin{acknowledgements}
The simulations were performed using the supercomputers hosted by the
CSC -- IT Center for Science Ltd.\ in Espoo, Finland, which is
administered by the Finnish Ministry of Education and in the HLRS
supercomputing centre in Stuttgart, Germany through the PRACE
allocation `SOLDYN'. Financial support
from the Academy of Finland ReSoLVE Center of Excellence (grant
No.\ 272157; MJK, NO) and grants No.\ 136189, 140970, 272786 (PJK),
the Swedish
Research Council grants 621-2011-5076 and 2012-5797 (AB),
and from the Estonian Research Council (Grant IUT40-1; JP)
are acknowledged,
as well as the Max-Planck/Princeton Center for Plasma Physics and
funding from the People Programme (Marie Curie Actions) of the
European Union's Seventh Framework Programme (FP7/2007-2013) under REA
grant agreement No.\ 623609 (JW).
We thank the anonymous referee for useful comments that helped us to
improve the contents and presentation of the manuscript.
\end{acknowledgements}

\bibliographystyle{aa}
\bibliography{paper}

\begin{appendix}

\section{Empirical Mode Decomposition} \label{subsect:EMD}

Fourier spectral analysis provides a general method for examining the
global energy--frequency distributions. For its validity, some crucial
conditions must hold: the system must be linear and the data must be
strictly periodic or stationary.  If these restrictions are not
satisfied, the non-stationary nature of the data causes the energy to
be spread over a wide frequency range. Additionally, deformed
wave-profiles, that are a direct consequence of nonlinear effects,
need additional harmonic components to be fitted. As a result, the
energy--frequency distribution can be misleading and hard to interpret
\citep{Huang98}.

There are different analysis methods developed for the purpose of
describing non-stationary signals. Well-known examples are short-time
Fourier transform and wavelet transform
\citep[e.g.][]{Qian02,Cohen95}. An alternative to the aforementioned
methods is the Hilbert-Huang transform (HHT) \citep{Huang98}. In
contrast to almost all the other methods, HHT works directly in the
temporal domain rather than in the corresponding frequency space and
the basis functions, also known as intrinsic mode functions (IMFs),
are derived from the data not selected \textit{a priori}.  The
decomposition makes implicitly the simple assumption that, at any
given time, the data may have many coexisting simple oscillatory modes
of significantly different frequencies, one superimposed on the other
\citep{Huang08}.

HHT is performed in two steps: firstly, using an algorithm called
Empirical Mode Decomposition (EMD), by which the signal is separated
into a set of IMFs; secondly, for each extracted mode, Hilbert
spectral analysis is applied which allows to describe each mode as an
analytical signal having the form $A(t)exp(i\varphi(t))$, where $A(t)$
is an instantaneous amplitude and $\varphi(t)$ an instantaneous phase.
For signals that result in such a form, the low frequency content is
in the amplitude term and the high frequency content in the
exponential term \citep{Cohen95}.  Having obtained a decomposition
into IMFs which satisfy the analytic signal conditions, we can
localize any event on the time as well as the frequency axis.  If
local time-dependent aspects of the IMFs are not of interest the
second step in the above algorithm may be replaced by a crude approach
that estimates the average mode period and amplitude by counting
zero-crossings and determining points of extrema.

One of the major drawbacks of the original EMD was the problem known
as mode mixing, which is a consequence of signal intermittency. To
overcome this problem, a noise-assisted data analysis method called
Ensemble EMD (EEMD) was proposed \citep{WU09}.  In \citet{Flandrin} it
was shown that when applied to Gaussian white noise, EMD acts as a
dyadic filter bank.  Utilizing this property allows to extract robust
and statistically significant IMFs.

EEMD has previously been applied to time series of different solar
activity proxies.  An example of the application to total solar
irradiance and sunspot data can be found in \citet{Barnhart11}.

\begin{figure*}[t!]
\begin{center}
  \includegraphics[width=0.95\textwidth]{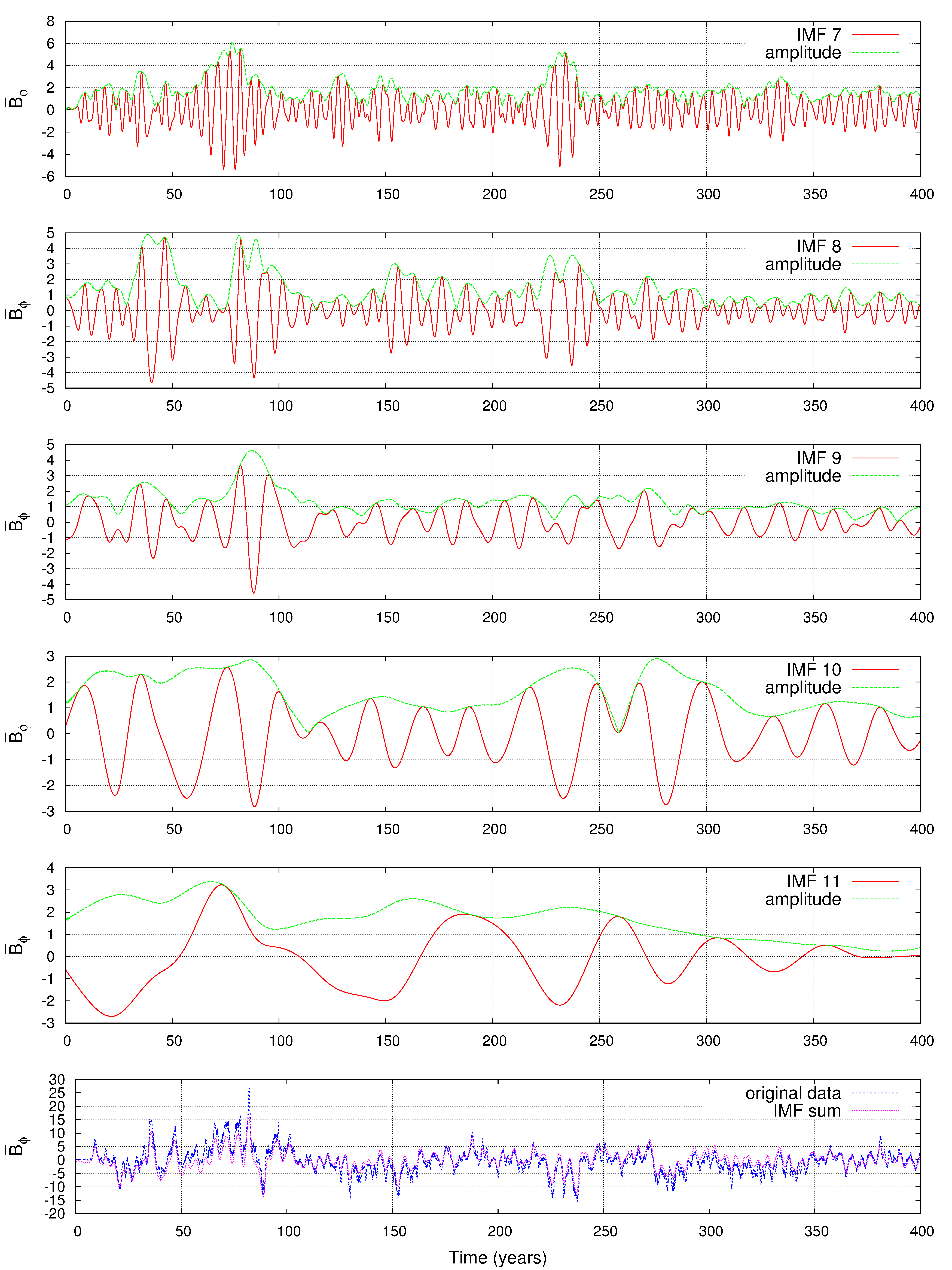}
\end{center}\caption[]{Modes 7--11 of $\mean{B}_{\phi}$ at $\Rb$ and latitude of $22^\circ$}.
\label{bphi_bot_modes}
\end{figure*}

As an illustration of how EEMD works we have selected a time series of
$\mean{B}_{\phi}$ taken at $\Rb$ and latitude of $22^\circ$.  This
time series was decomposed into 12 IMFs, most of the energy being
distributed over modes 7--11, each of which individually contribute
more than 10\% to the full energy.  The given modes, with the
corresponding instantaneous amplitudes are shown on the top 5 panels
in Fig.~\ref{bphi_bot_modes}. On the bottom panel the original time
series of $\mean{B}_{\phi}$ as well as the sum of these five modes are
depicted.  The small difference between the curves comes from the fact
that the less significant modes (1--6, 12 and 13) have been excluded
from the sum. We also note that the form of the IMFs is not precisely
satisfying the criteria of an analytic signal on one hand due to the
finite precision limit in the IMF extraction algorithm and on the
other hand due to the relatively small ensemble size (the noise does
not fully cancel out).

\section{$D^2$ phase dispersion statistic} \label{subsect:D2}

The $D^2$ phase dispersion statistic was first introduced in
  \citet{Pelt83}. Recent applications of the method can be found in
  \citet{LMOPHHJS13}, \cite{KKKBOP15}, and \cite{Olspert15}.
  Similarly to EEMD, the $D^2$ statistic is well suited for
  non-stationary data.  However, while EEMD aims at decomposing the
  initial time series into a set of quasi-harmonic signals, the
  purpose of $D^2$ is to find a set of average cycle lengths, $P_{\rm
    m}$, which are consistent with the full data set. The $D^2$ statistic
  is defined as follows:
\begin{equation}
D^2(P,\Delta t) = \frac{\sum\limits_{i = 1}^{N - 1} {\sum\limits_{j = i + 1}^N {g(t_i } } ,t_j ,P,\Delta t)[f(t_i ) - f(t_j )]^2 }{2\sigma^2\sum\limits_{i = 1}^{N - 1} {\sum\limits_{j = i + 1}^N {g(t_i } } ,t_j ,P,\Delta t)},
\end{equation}
where $f(t_i),i = 1,\dots,N$ is the input time series, $\sigma^2$ is its variance, $g(t_i,t_j,P,\Delta t)$ is the selection function, which is significantly greater than zero only when
\begin{eqnarray}
t_j  - t_i  &\approx& kP,k =  \pm 1, \pm 2, \ldots {\rm \ \ \ and}\\
\left| {t_j  - t_i } \right| &\lessapprox& \Delta t,
\end{eqnarray}
where $P$ is the trial period and $\Delta t$ is the so-called
coherence time, which is the measure of the width of the sliding time window
wherein the data points are taken into account by the statistic.
More precisely, in the given study the function $g$ has been chosen as
the product of cosine and Gaussian functions. Here the former is
depending only on the phase difference of given points w.r.t given
trial period and the latter only on the time difference between
the corresponding points:
\begin{eqnarray}
g(t_i,t_j,P,\Delta t) &=& g_1(t_i,t_j,P)\cdot g_2(t_i,t_j,\Delta t),\\
g_1(t_i,t_j,P) &=& \frac{1}{2}\left(\cos\left(2\pi \cdot \rm{frac}\left(\frac{t_j-t_i}{P}\right)\right)+1\right ),\\
g_2(t_i,t_j,\Delta t) &=& \exp\left(-\ln 2 \left( \frac{t_j-t_i}{\Delta t}\right )^2\right ).
\end{eqnarray}

  As $\Delta t$ is made
  shorter, we match nearby cycles in a progressively narrower region,
  and consequently, estimate a certain mean cycle length $P_{\rm m}$,
  not necessarily coherent for the full time span.
  At large values of $\Delta t$, i.e. when the time window width approaches
  the time span of the data, the $D^2$ spectrum approaches
  the result that would be obtained using e.g. Fourier transform.
  More precisely,
  $P_{\rm m}$ is determined from the dispersion spectrum calculated
  for the longest coherence time for which the shape of the minimum
  is still symmetric and singular (with a single local minimum within
  reasonably narrow search region to exclude possible minima from higher harmonics).
  Denoting this optimal coherence time by
  $\Delta t_{\rm opt}$ we define the mean cycle length as:
  \begin{equation}
  P_{\rm m}=\underset{P}{\arg \min} \{D^2(P, \Delta t_{\rm opt}) \}.
  \end{equation}
  When visualizing the spectrum, we plot the cycle length $P$ on vertical
  axis, but instead of directly plotting the coherence time on
  horizontal axis, we use $\Delta t/P$ instead. This is just a count
  of cycles with a corresponding period that fit into the
  time interval equal to given coherence time. We call this entity coherence length.

\end{appendix}
\end{document}